\documentclass[aps,prd,floatfix,showpacs,10pt,nofootinbib,twocolumn]{revtex4-1}
\usepackage{amsmath}
\usepackage{amsfonts}
\usepackage{amssymb}
\usepackage{amsthm}
\usepackage{mathrsfs}
\usepackage{multirow}
\usepackage{color}
\usepackage{array}
\usepackage{tikz}
\usepackage{amsmath,amssymb}
\usetikzlibrary{calc}
\usetikzlibrary{decorations}
\usetikzlibrary{decorations.fractals}
\usetikzlibrary{decorations.pathreplacing}
\usetikzlibrary{decorations.pathmorphing}
\usetikzlibrary{decorations.markings}
\usetikzlibrary{positioning,matrix}
\usetikzlibrary{shapes.arrows}
\usetikzlibrary{shapes.symbols}
\usetikzlibrary{shapes.misc}
\usetikzlibrary{shapes.geometric}
\usetikzlibrary{through}
\usetikzlibrary{mindmap,backgrounds}
\usetikzlibrary{fit}
\usetikzlibrary{arrows,positioning}
\usetikzlibrary{external}
\usepackage[colorlinks, urlcolor=black, citecolor=blue, link
color=black]{hyperref}

\def\nn    {\nonumber}

\def\beq{\begin{equation}}
\def\eeq{\end{equation}}
\def\beqa{\begin{eqnarray}}
\def\eeqa{\end{eqnarray}}
\def\L{{\mathcal L}}
\def\eff{{\rm eff}}
\def\plusX{+X}

\allowdisplaybreaks

\begin{document}
\title{
Identifying a $Z'$ behind $b \to s \ell \ell$ anomalies at the LHC
}

\author{M. Kohda}
\author{T. Modak}
\affiliation{Department of Physics, National Taiwan University, Taipei 10617, Taiwan}
\author{A.~Soffer}
\affiliation{Tel Aviv University, Tel Aviv, 69978, Israel}

%%%%%%%%%%%%%%%%%%%%%%%%%%%%%%%%%%%%%%%%%%%%%%%%%%%%%%%%%%%%%%%%%%%%%%%%%

%%%%%%%%%%%%%%%%%%%%%%%%%%%%%%%%%%%%%%%%%%%%%%%%%%%%%%%%%%%%%%%%%%%%%%%%%

\begin{abstract}
Recent $b\to s\ell\ell$ anomalies may imply the existence of a new $Z'$ boson 
with left-handed $Z'bs$ and $Z'\mu\mu$ couplings.
Such a $Z'$ may be directly observed at LHC via
$b \bar s \to Z' \to \mu^+\mu^-$, and its relevance to $b\to s\ell\ell$
may be studied by searching for the process $gs \to Z'b \to \mu^+\mu^- b$.
In this paper, we analyze the capability of the 14 TeV LHC to observe the $Z'$
in the $\mu^+ \mu^-$ and $\mu^+\mu^- b$ modes based on an effective model 
with major phenomenological constraints imposed.
We find that both modes can be discovered with 3000 fb$^{-1}$ data
if the $Z'bs$ coupling saturates the latest $B_s-\bar B_s$ mixing limit from UTfit
at around $2\sigma$.
Besides, a tiny right-handed $Z'bs$ coupling, if it exists, 
opens up the possibility of a relatively large left-handed counterpart,
due to cancellation in the $B_s-\bar B_s$ mixing amplitude.
In this case, we show that even a data sample of 
$\mathcal{O}(100)$ fb$^{-1}$ would enable discovery of both modes.
We further study the impact of a $Z'bb$ coupling as large as the $Z'bs$ coupling.
This scenario enables discovery of the $Z'$ in both modes
with milder effects on the $B_s-\bar B_s$ mixing,
but obscures the relevance of the $Z'$ to $b \to s\ell\ell$.
Discrimination between the $Z'bs$ and $Z'bb$ couplings may come from
the production cross section for the $Z'b\bar{b}$ final state.
However, we do not find the prospect for this to be promising.
\end{abstract}

\maketitle

\section{Introduction}

Bottom-quark transitions of $b\to s$ have been of great interest as
a means for studying physics beyond the standard model (SM) since 
the observation of the decay $B\to K^* \gamma$ by the CLEO
collaboration~\cite{Ammar:1993sh}.  Involving a flavor-changing
neutral current (FCNC), such processes are possible in the SM
only at loop level, providing unique sensitivity to new physics (NP).

The high production rate and detection efficiency for bottom hadrons
at the LHCb experiment have enabled precision tests that probe physics
at high energy scales. Based on the Run-1 data, 
LHCb measurements of several observables related to 
$b\to s \ell^+ \ell^-$ ($\ell = e$ or $\mu$) transitions are in tension with
SM predictions. The most notable discrepancies are found in the
$P_5'$~\cite{DescotesGenon:2012zf} angular-distribution observable 
for the $B^0\to K^{*0}\mu^+\mu^-$ decay~\cite{Aaij:2013qta,Aaij:2015oid}, and
in the lepton-flavor universality observables 
$R_K\equiv \mathcal{B}(B^+ \to K^+ \mu^+\mu^-)/\mathcal{B}(B^+ \to K^+ e^+ e^-)$~\cite{Aaij:2014ora} and $R_{K^*}\equiv\mathcal{B}(B^0 \to
K^{*0} \mu^+ \mu^-)/\mathcal{B}(B^0 \to K^{*0} e^+ e^-)$~\cite{Aaij:2017vbb}.
Moreover, measured differential branching ratios for exclusive $b\to s\mu^+\mu^-$
decays such as $B^{0} \to K^{0} \mu^+\mu^-$, $B^+ \to K^+ \mu^+\mu^-$,
$B^+\to K^{*+}\mu^+\mu^-$~\cite{Aaij:2014pli}, 
$B^0 \to K^{*0}\mu^+\mu^-$~\cite{Aaij:2016flj}, 
$B_s^0 \to \phi\mu^+\mu^-$~\cite{Aaij:2015esa}
and $\Lambda_b^0 \to \Lambda \mu^+ \mu^-$~\cite{Aaij:2015xza}
are consistently lower than SM predictions in the dimuon-invariant mass range below 
the $J/\psi$ threshold. 
ATLAS and CMS are also capable of studying $b \to s \mu^+\mu^-$ transitions.
They have performed angular analyses for $B^0\to K^{*0}\mu^+\mu^-$ 
with 8 TeV data~\cite{ATLAS:2017dlm,Sirunyan:2017dhj},
where the measured $P_5'$ by ATLAS supports the discrepancy found by LHCb
while the measurements by CMS are in agreement with SM predictions.
Belle~\cite{Wehle:2016yoi} reports an angular analysis for $B \to K^* \ell^+\ell^-$,
finding mild tension in $P_5'$ in the muon mode, but not in the electron mode.
The measurements will be significantly improved with more data collected by LHC,
as well as with the upcoming Belle II experiment~\cite{Abe:2010gxa}.

While the statistical significance of each discrepancy is not large enough,
there is excitement about the possibility that their combination might
suggest the presence of NP.
To investigate this possibility, global-fit analyses based on 
the effective Hamiltonian formalism have been performed by several groups
(see Refs.~\cite{Capdevila:2017bsm,Altmannshofer:2017yso,DAmico:2017mtc,
  Hiller:2017bzc,Geng:2017svp,Ciuchini:2017mik,Celis:2017doq,Hurth:2017hxg}
for studies that include the recent $R_{K^*}$~\cite{Aaij:2017vbb} result).
These fits find that the tensions in the $b \to s \ell\ell$
observables can be simultaneously alleviated to a great extent by a NP
contribution in a single Wilson coefficient. Most authors find this to
be the coupling between the left-handed (LH) $b\to s$ current and
either the LH or vector muon current.\footnote{However, we note the
  existence of a fit to other $B\to K^* \mu^+ \mu^-$
  observables~\cite{Karan:2016wvu}, which indicates a NP contribution
  in the right-handed $b\to s$ current.}

In particular, the findings of the global-fit analyses motivate
studying a new gauge boson, $Z'$, with FCNC interactions.  Many
phenomenological studies of such a $Z'$ have been performed (see, e.g.,~\cite{
  Altmannshofer:2013foa, Gauld:2013qba, Buras:2013qja, Gauld:2013qja,
  Buras:2013dea, Ko:2013zsa, Ahmed:2014vqa, Altmannshofer:2014cfa,
  Bhattacharya:2014wla, Crivellin:2015mga, Crivellin:2015lwa,
  Niehoff:2015bfa, Sierra:2015fma, Altmannshofer:2015sma,
  Crivellin:2015era, Celis:2015ara, Belanger:2015nma,
  Falkowski:2015zwa, Descotes-Genon:2015uva, Allanach:2015gkd,
  Buras:2015kwd, Fuyuto:2015gmk, Chiang:2016qov, Kim:2016bdu,
  Altmannshofer:2016oaq, Hisano:2016afc, Altmannshofer:2016brv,
  Ko:2017quv, Bhatia:2017tgo, Hou:2017ozb, Alok:2017jgr, Greljo:2017vvb, Alok:2017sui,
  Bonilla:2017lsq, Ellis:2017nrp, Bishara:2017pje, Alonso:2017uky,
  Chiang:2017hlj, King:2017anf, Chivukula:2017qsi, Cline:2017ihf,
  Cox:2017eme,Baek:2017sew, Romao:2017qnu, Cox:2017rgn,Faisel:2017glo,
  Dalchenko:2017shg, Allanach:2017bta,DiChiara:2017cjq,Choudhury:2017ijp,
  Antusch:2017tud,Fuyuto:2017sys,Raby:2017igl, DiLuzio:2017fdq, Chala:2018igk}).

Produced at LHC via $b\bar{s} \to Z'$ and undergoing the decay $Z' \to
\mu^+\mu^-$, the $Z'$ may be discovered in dimuon resonance
searches. Such a discovery by itself, however, would not reveal the
relevance of the observed resonance to the $b \to s\ell\ell$
anomalies.  Comparison with searches in the electron mode can test
lepton-flavor universality in the $Z'$ couplings, but one should also establish the
coupling to the $b\to s$ current.  
In principle, this can be done with the decay $Z' \to b\bar s$.
However, this decay is suppressed relative to $Z' \to \mu^+\mu^-$ due
to the $B_s-\bar B_s$ mixing constraint, as we discuss below,
and its detection suffers from overwhelming QCD background.
On the
other hand, the $Z'bs$ coupling can induce the process $gs \to
Z'b$. Therefore, this coupling can be explored through production
modes of the $Z'$ at LHC.

In this paper, we investigate the prospects for direct observation of the $Z'$, 
as well as determination of the flavor structure of its couplings at LHC, with 
$\sqrt{s}=14$~TeV.
We argue that achieving this dual goal requires measuring
the cross sections of both $pp\to Z'\plusX$ and $pp\to Z'b\plusX$,
where $X$ refers to additional activity in the $pp$ collision.
We employ an effective-model description of the $Z'$ couplings, 
and assume that it is the source of all
the NP required to alleviate the tensions in $b\to s\mu^+\mu^-$.
We focus mainly on the role of the $Z'bs$ coupling in the $Z'$ production processes 
$pp\to Z'\plusX$ and $pp\to Z'b\plusX$. 
The constraints on the leptonic coupling of the $Z'$ are much weaker than those
on the $Z'bs$ coupling. Therefore, we use 
$Z' \to \mu^+\mu^-$ as the main discovery mode.
As the LH $Z'bs$ coupling is accompanied by a LH $Z'bb$ coupling 
in many UV complete models (see, e.g.
Refs.~\cite{Ko:2017quv,Bonilla:2017lsq,Alonso:2017uky,Altmannshofer:2014cfa}), 
we also study scenarios in which the $Z'bb$ coupling is of the same order as 
the $Z'bs$ coupling.
We note that a larger $Z'bb$ coupling implies larger cross sections and easier 
discovery of the $Z'$ at LHC, but obscures the role of the $Z'$ in $b \to s\ell\ell$.
Therefore, this case is not the focus of our work.
We also consider the process $pp\to Z'b\bar{b}\plusX$ for discrimination between
the $Z'bs$ and $Z'bb$ couplings, where the latter coupling uniquely contributes 
via $gg \to Z'b\bar{b}$.

We emphasize that our purpose is different from that of existing
studies, which
focus on discovery and/or constraint of the $Z'$ rather than
testing its role in $b \to s\ell\ell$.
Recent studies on the impact of existing dimuon resonance searches 
on a $Z'$ motivated by the $b \to s\ell\ell$ anomalies can be found, e.g., 
in Refs.~\cite{Chivukula:2017qsi,Dalchenko:2017shg}.
Ref.~\cite{Dalchenko:2017shg} also studies the future sensitivity at LHC,
exploiting the use of additional $b$ jets 
for background suppression.
However, it targets a $Z'bb$ coupling that is much larger than the $Z'bs$ coupling.
The future sensitivities at LHC and a 100~TeV $pp$ collider are studied
in Ref.~\cite{Allanach:2017bta} with an extrapolation of existing ATLAS limit.

We will show that the cross sections for  $b\bar s \to Z'$ and $gs \to Z'b$ are 
limited by a rather tight constraint on the $Z'bs$ coupling, 
which originates from the $B_s-\bar B_s$ mixing.
This severely restricts the discovery potential of the $Z'$, 
unless the $Z'bb$ coupling 
is comparable or larger than the $Z'bs$ coupling.
As the $B_s-\bar B_s$ mixing constraint is indirect, the actual limit on the $Z'bs$ coupling
depends on the details of the UV-complete model.
In particular, a nonzero right-handed (RH) $Z'bs$ coupling can drastically change 
the constraint, due to the large $(V-A)\otimes (V+A)$
term~\cite{Buras:2001ra} in the $B_s-\bar B_s$ mixing amplitude.
Although there is no strong indication of a RH $b\to s$ current in the majority 
of the global-fit analyses, even a tiny RH $Z'bs$
coupling would allow for a large LH $Z'bs$ coupling due to the cancellation.
This would significantly boost the $Z'$ production cross sections.
We investigate the discovery potential in this case as well. 

This paper is organized as follows. In Sec.~\ref{sec:model}, we introduce 
the effective model for our collider study.
In Sec.~\ref{sec:pam}, we evaluate 
existing phenomenological constraints on the relevant couplings of the $Z'$ boson to
quarks and leptons. This is carried out for two representative $Z'$
mass values, $m_{Z'}=200$ and 500~GeV.
In Sec.~\ref{coll}, we study the signal and background cross sections for the three 
processes of interest, $pp\to Z'\plusX$, $Z'b\plusX$ and $Z'b\bar{b} \plusX$,
given the coupling constraints. We then proceed to estimate the signal
significances for the full integrated luminosity of the High-Luminosity LHC 
(HL-LHC) program, $\L=3000~{\rm fb}^{-1}$.
In Sec.~\ref{sec:RHb2s}, we discuss the impact of a tiny but nonzero RH $Z'bs$
coupling, which allows discovery with smaller integrated luminosities. 
Summary and discussions are given in Sec.~\ref{summ}.

%%%%%%%%%%%%%%%%%%%%%%%%%%%%%%%%%%%%%%%%%%%%%%%%%%%%%%%%
\section{Effective model}
\label{sec:model}

We describe the $Z'$ couplings to the SM fermions with the effective Lagrangian 
\begin{align}
\mathcal{L} \supset 
 &-  Z'_\alpha \big[ g^L_{bb}~\bar{b}\gamma^\alpha P_L b 
 +g^L_{bs} \left( \bar{b}\gamma^\alpha P_L s + \bar{s}\gamma^\alpha P_L b \right)\nn\\
 &+g^{L}_{\mu\mu} \left( \bar{\mu}\gamma^\alpha P_L \mu 
   +\bar{\nu}_\mu \gamma^\alpha P_L \nu_\mu\right)
 +g^{R}_{\mu\mu}~\bar{\mu}\gamma^\alpha P_R \mu \big],\label{efflag}
\end{align}
where $P_{L,R}= (1\mp \gamma_5)/2$, and
$g^L_{bb}$, $g^L_{bs}$ and $g^{L,R}_{\mu \mu}$ are coupling constants.
For simplicity, 
since no significant $CP$ violation has been observed in the relevant observables, 
we take the couplings to be real.
In addition to the LH $Z'bs$ coupling and LH and RH $Z'\mu\mu$
couplings motivated by the $b\to s \ell^+\ell^-$ global fits,
we introduce the LH $Z'bb$ coupling predicted in many UV complete models.
We take $g^{L}_{\mu \mu}$ to also be the coupling to the muon neutrino, as required by
the SU(2)$_L$ gauge symmetry, and since observables containing neutrinos give
meaningful constraints on the $Z'$ parameters, as shown below.
The RH $Z'bs$ coupling is not included at this stage, as it is
discussed only in Sec.~\ref{sec:RHb2s}.

Since we take $m_{Z'} \gg m_b$, we integrate out the $Z'$ to
obtain its contributions to the effective Hamiltonian for $b \to s
\mu^+ \mu^-$ transitions, given by
\begin{align}
\Delta\mathcal{H}_{\eff} =& {\mathcal{N}} \big[
 C_9^{\rm NP} (\bar{s}\gamma^\alpha P_L b)(\bar{\mu}\gamma_\alpha \mu)\nn\\
  &\quad+C_{10}^{\rm NP} (\bar{s}\gamma^\alpha P_L b) (\bar{\mu}\gamma_\alpha \gamma_5\mu)\big ] +{\rm h.c.}, \label{eq:bsll}
\end{align}
where $\mathcal{N}=-\frac{\alpha G_F}{\sqrt{2}\pi} V_{tb}V^*_{ts}$, and
\begin{align}
C_9^{\rm NP}=  \frac{g^{L}_{bs}~g^{V}_{\mu\mu}}{\mathcal{N} m_{Z'}^2}, \quad
C_{10}^{\rm NP}=  \frac{g^{L}_{bs}~g^{A}_{\mu\mu}}{\mathcal{N} m_{Z'}^2}, \label{c9-10-Zp}  
\end{align}
are the Wilson coefficients, 
with the vector and axial-vector muon couplings defined by
\begin{align}
g^{V}_{\mu\mu} \equiv \frac{ g_{\mu\mu}^R +g_{\mu\mu}^L }{2}, \quad
g^{A}_{\mu\mu} \equiv \frac{ g_{\mu\mu}^R -g_{\mu\mu}^L }{2}.
\end{align}
With the parametrization of the Cabibbo-Kobayashi-Maskawa (CKM) matrix
in Ref.~\cite{Olive:2016xmw}, $C_{9, 10}^{\rm NP}$ are treated as
real to a good approximation.

Motivated by the global-fit analyses, we consider two possibilities
for the chiral structure of the muon couplings: a vector coupling and
a LH coupling.
For each of these, we extract constraints on the
couplings from global-fit analyses presented in
Ref.~\cite{Capdevila:2017bsm}.
The first analysis uses all available $b\to s\ell\ell$ data from LHCb,
ATLAS, CMS and Belle. This yields the following two scenarios:
\begin{enumerate}
\item[(i)] Vector coupling ($g_{\mu\mu}^L =
  g_{\mu\mu}^R$) and, hence, $C_{10}^{\rm NP}=0$.
In this case, the best fit value is ${\rm Re}~C^{\rm NP}_{9} = -1.11$,
with a 2-standard-deviation ($2\sigma$) range of
\begin{align}
-1.45 \leq {\rm Re}~C_9^{\rm NP} \leq -0.75.
\label{eq:WC-fit-i}
\end{align}
\item[(ii)] LH coupling ($g_{\mu\mu}^R = 0$),
so that  $C_{9}^{\rm NP} = -C_{10}^{\rm NP}$.
For this scenario, the best fit value is ${\rm Re}~C^{\rm NP}_{9} =
-{\rm Re}~C^{\rm NP}_{10}=-0.62$, with a $2\sigma$ range
\begin{align}
-0.88 \leq {\rm Re}~C_9^{\rm NP} = -{\rm Re}~C_{10}^{\rm NP} \leq -0.37.
\label{eq:WC-fit-ii}
\end{align}
\end{enumerate}

The authors of Ref.~\cite{Capdevila:2017bsm} also present results when
taking into account only lepton-flavor universality observables, such
as $R_{K^{(*)}}$ measured by LHCb~\cite{Aaij:2014ora, Aaij:2017vbb}
and the differences $Q_4$ and $Q_5$~\cite{Capdevila:2016ivx} between
angular observables in $B\to K^*\mu^+\mu^-$ and $B\to K^* e^+e^-$,
measured by Belle~\cite{Wehle:2016yoi}. 
This leads to the following two scenarios:
\begin{enumerate}
\item[(i')] In the case of vector coupling,
the best-fit value is ${\rm Re}~C^{\rm NP}_{9} = -1.76$, with $ -3.04\leq {\rm Re}~C^{\rm NP}_{9} \leq -0.76$ at $2\sigma$
interval. 

\item[(ii')] For the LH-coupling case, the best-fit value is found to be ${\rm Re}~C^{\rm
  NP}_{9} = -{\rm Re}~C^{\rm NP}_{10} =-0.66$, while at $2\sigma$
range  $-1.04\leq{\rm Re}~C^{\rm NP}_{9} = -{\rm Re}~C^{\rm NP}_{10}\leq -0.32$.
\end{enumerate}

In the rest of the paper we explore scenario~(i) in detail, and
occasionally comment on differences with respect to the other
scenarios. Generally, these differences are small and do not
significantly affect our main results in Section~\ref{coll}.

%%%%%%%%%%%%%%%%%%%%%%%%%%%%%%%%%%%%%%%%%%%%%%%%%%%%%%%%%%%%%%%
\section{Allowed parameter space}
\label{sec:pam}

\begin{figure*}[tb!]
\centering
\includegraphics[width=.4 \textwidth]{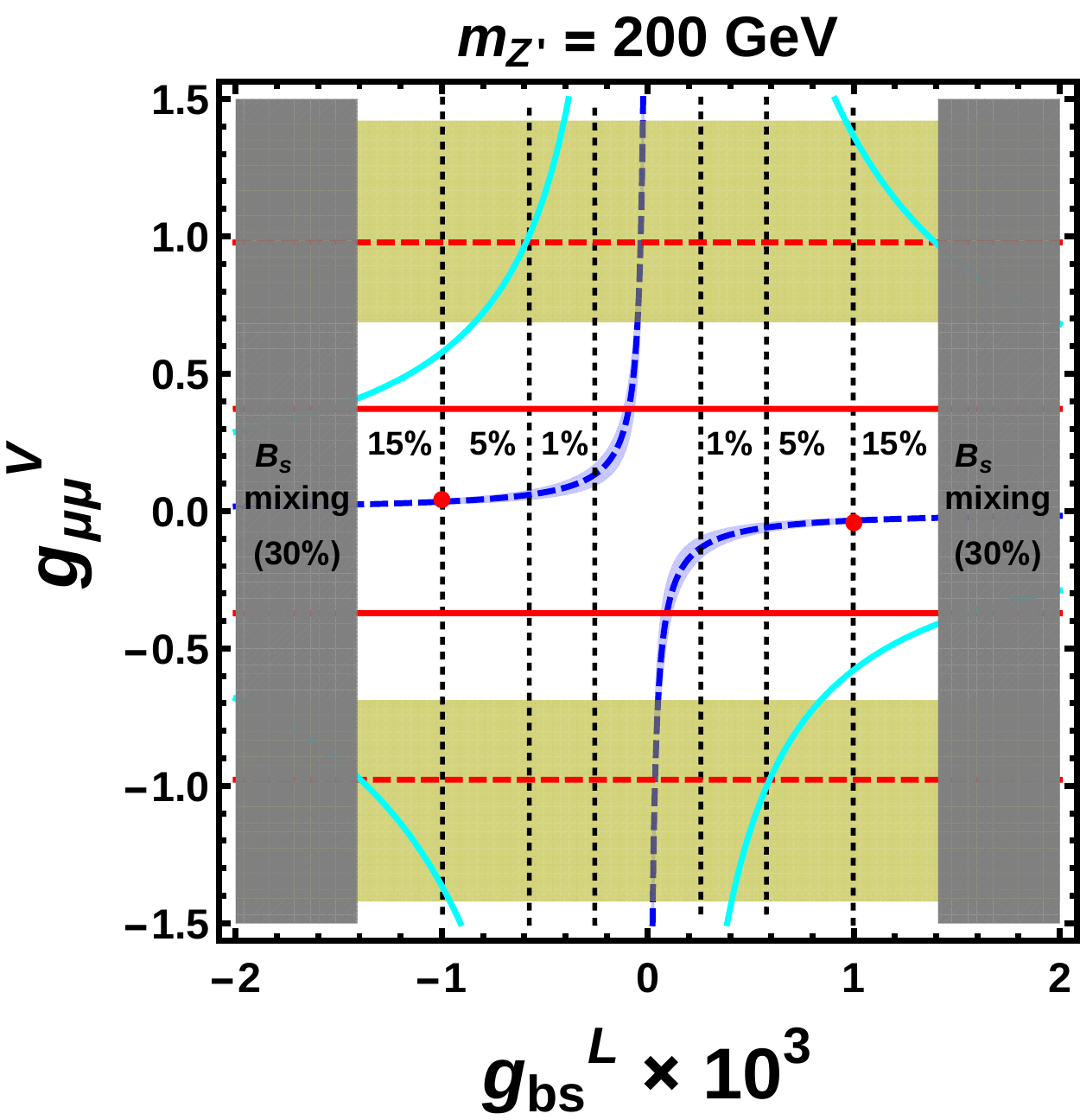}
\includegraphics[width=.58\textwidth]{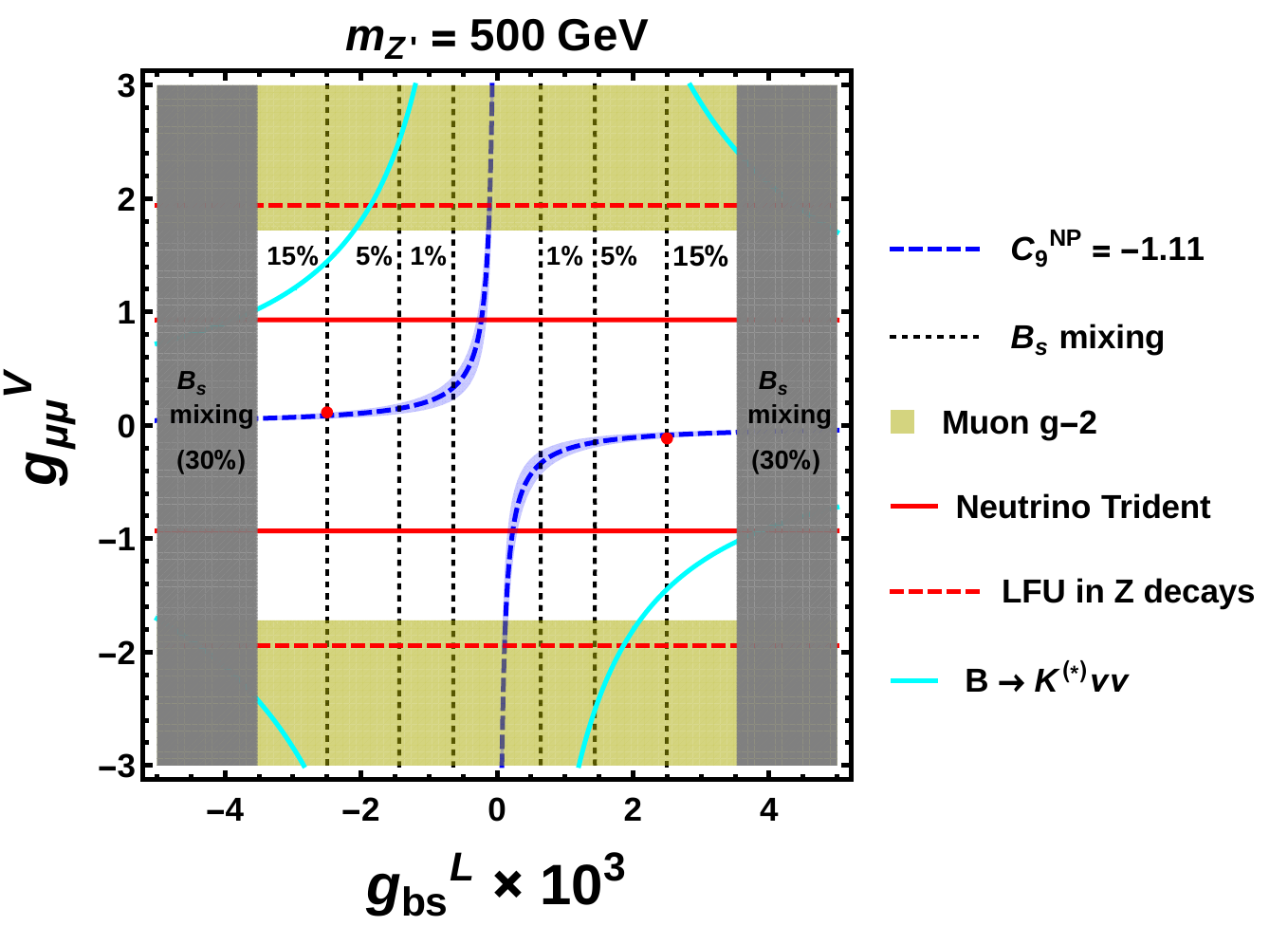}
\caption{
Constraints on the vector $Z'\mu\mu$ coupling 
$g^V_{\mu\mu} = g^L_{\mu\mu} = g^R_{\mu\mu}$ vs.
the LH $Z'bs$ coupling $g^L_{bs}$ (in units of $10^{-3}$) 
in scenario~(i), 
for $m_{Z'}=200$~GeV [left] and $500$~GeV [right]. 
The red dots show the benchmark points discussed in Section~\ref{coll}.
See the main text for details.
}
\label{const-V-mu}
\end{figure*}

In Fig.~\ref{const-V-mu} we show the various constraints on $g^{L}_{bs}$
vs. $g^{V}_{\mu\mu}$ in scenario~(i), for the representative
$Z'$-mass values of $m_{Z'}=200$ and 500~GeV. 
The relevant inputs and constraint calculation methods
are described in the remainder of this section.

Before embarking on this detailed description, we note that, as seen
in Fig.~\ref{const-V-mu}, the limits on $g^V_{\mu\mu}$ are much weaker
than those on $g^L_{bs}$.  Therefore, the $Z'$ is likely to decay
primarily to leptons, so that its decays into quarks can be ignored.
The leptonic-decay dominance simplifies the discussion. In fact, it is
also essential for direct observation of the $Z'$ at LHC, since
searches with $Z' \to b\bar b, b\bar s$ suffer from overwhelming QCD
background.  
Thus, in the scenario~(i), where $g_{\mu\mu}^L = g_{\mu\mu}^R \neq 0$,
the dominant branching ratios are
\begin{align}
\mathcal{B}(Z'\to\mu^+\mu^-) \simeq \frac{2}{3},  
\mathcal{B}(Z'\to\nu_\mu\bar\nu_\mu) \simeq \frac{1}{3}~.
\label{brs}
\end{align}
In scenario~(ii),  each of these branching ratios is 50\%.

%%%%%%%%%%%%%
%\subsection{$b\to s\ell\ell$ Global fit}
%\label{sec:pam:bsll}

To incorporate the results of $b\to s\ell\ell$ global fits,
we use Eq.~(\ref{c9-10-Zp}) to convert the $2\sigma$ constraints of
Eq.~(\ref{eq:WC-fit-i}) to the space of $g^{L}_{bs}$
vs. $g^{V}_{\mu\mu}$. The result is given by the blue hyperbolae in
Fig.~\ref{const-V-mu}

As an example of the impact of the choice of scenario among those
listed in Sec.~\ref{sec:model}, we note that in scenario~(i'), the
resulting values of $g^V_{\mu\mu}$ are generically higher than those
shown in Fig.~\ref{const-V-mu}, and have a wider range.  
Although this somewhat changes the $Z'\to \mu^+\mu^-$ branching ratio,
Eq.~(\ref{brs}) is still satisfied. Therefore, the results we obtain
in Section~\ref{coll} are not affected.

%%%%%%%%%%%% 
The most important constraint on the $Z'bs$ coupling comes from the $B_s$ mixing.
With the tree-level $Z'$ exchange contribution,
the total $B_s - \bar B_s$ mixing amplitude 
relative to the SM one (see, e.g. Ref.~\cite{Lenz:2010gu}) is given by
\begin{align}
 \frac{M_{12}}{M_{12}^{\text{SM}}}=1+\frac{\left({g^L_{bs}}\right)^2}{m_{Z'}^2}
 \left(\frac{g^2_2}{16\pi^2 v^2}(V_{tb} V_{ts}^*)^2 S_0\right)^{-1}, \label{eq:M12}
\end{align}
where $S_0\simeq2.3$ is an Inami-Lim function, $g_2$ is the SU(2)$_L$ gauge coupling,
$v \simeq 246$~GeV, and a common QCD correction factor is assumed for
the SM and NP contributions.
The mass difference $\Delta m_{B_s^0} = 2|M_{12}|$ is precisely measured,
at the per~mill level~\cite{Olive:2016xmw}, while the calculation of $M_{12}$ suffers from
various sources of uncertainty.
One of the dominant uncertainties is the CKM factor, with an uncertainty of
$\sim 5$\%~\cite{Charles:2015gya,Bona:2006sa}.
The other is the hadronic matrix element, obtained from the lattice. 
The average of $N_f = 2+1$ lattice results 
compiled by FLAG in 2016~\cite{Aoki:2016frl} implied a $\sim 12$\% uncertainty in $M_{12}^{\rm SM}$.
Recently, the situation was greatly improved with the advent of the accurate estimate 
by the Fermilab Lattice and MILC collaborations~\cite{Bazavov:2016nty},
which pushes down the uncertainty of  the FLAG average to $\sim 6$\%.
(See December 2017 update on the FLAG website~\cite{Aoki:2016frl}.)
A global analysis of the CKM parameters by CKMfitter~\cite{Charles:2015gya},
published before the recent lattice result~\cite{Bazavov:2016nty},
gives a constraint on $M_{12}$ with NP as $|M_{12}/M_{12}^{\rm SM}| = 1.05^{+0.14}_{-0.13}$. 
On the other hand, the Summer 2016 result~\cite{Bona:2006sa} by UTfit,
which includes the result of Ref.~\cite{Bazavov:2016nty}, 
constrains NP with a better precision: $|M_{12}/M_{12}^{\rm SM}| = 1.070 \pm 0.088$.
As we assume $g^L_{bs}$ to be real, the $Z'$ contribution always enhances 
$|M_{12}|$.
If one takes these uncertainties to be Gaussian, these results imply
$|M_{12}/M_{12}^{\rm SM}| < 1.32$~\cite{Charles:2015gya} or $1.25$ \cite{Bona:2006sa}
at $2\sigma$ for the CKMfitter and UTfit results, respectively.
In this paper, we explore NP contributions to $|M_{12}/M_{12}^{\rm SM}|$
at the level of up to $30\%$.
Excluding larger contributions leads to the gray-shaded regions in Fig.~\ref{const-V-mu}.
Future improvements in lattice calculations and measurements of the CKM parameters 
would tighten the constraint~\cite{Bona-ICHEP2016}.
In Fig.~\ref{const-V-mu}, we illustrate the impact of possible future improvements 
by the vertical dotted lines for deviation of $M_{12}$ from SM by 15\% or 5\%.
%\AS{We need to motivate the 15, 5, 1\% lines.}

%
We note that while the CKMfitter~\cite{Charles:2015gya} and
UTfit~\cite{Bona:2006sa} results are tolerant to a NP contribution
that enhances $|M_{12}|$, there are studies that find the SM prediction of 
$\Delta m_{B_s^0} =2|M_{12}|$ to be larger than the measured value,
slightly favoring NP that reduces $|M_{12}|$.
In particular, a recent study~\cite{DiLuzio:2017fdq}, which adopts the 2017 FLAG
result~\cite{Aoki:2016frl}, finds the SM prediction to be $1.8\sigma$ above
the measured value. %resulting in a rather strong constraint on the $Z'$.
Their result can be read as
$|M_{12}/M_{12}^{\rm SM}| \simeq 0.89 \pm 0.06$,
which allows an enhancement by NP only up to $\sim 1$\% at 2$\sigma$.
The rather small uncertainty is in part due to a smaller uncertainty
of 2.1\% assigned to the CKM factor.
Addressing the discrepancy among the theoretical calculations 
is beyond the scope of this paper.
Instead, the 1\% vertical dotted lines are also shown in Fig.~\ref{const-V-mu}
for illustrating the impact of the result by Ref.~\cite{DiLuzio:2017fdq}.

%%%%%%%%%%%%%%%%%%%%%%%%%%%%%  

The nonzero LH $Z'\mu\mu$ coupling implies also the existence of a
$Z'\nu\nu$ coupling, due to SU(2)$_L$. Therefore, constraints are also
set by $B \to K^{(*)} \nu\bar{\nu}$.
The effective Hamiltonian for $b \to s \nu \bar\nu$ is~\cite{Buras:2014fpa}
\begin{align}
 \mathcal{H}_{eff}^{\nu} = \mathcal{N} \sum_{\ell = e,\mu, \tau} C_L^\ell
 \left(\bar{s}\gamma^\alpha P_L b\right)
 \left[ \bar{\nu}_\ell \gamma_\alpha(1-\gamma_5)\nu_\ell \right] +{\rm h.c.},
\end{align}
where $C_L^\mu = C_L^{\rm SM}+C_L^{\rm NP}$ and 
$C_L^\ell = C_L^{\rm SM}$ ($\ell = e, \tau$)
are lepton-flavor dependent Wilson coefficients.
The SM contribution is lepton-flavor universal and is given by
$C_L^{\rm SM} = -X_t/s_W^2$ with $X_t = 1.469\pm0.017$~\cite{Brod:2010hi}.
The $Z'$ contribution is given by
\begin{align}
 C_L^{\rm NP}= \frac{g^L_{bs}~g^L_{\mu\mu}}{2\mathcal{N}m_{Z'}^2}.
\end{align}
Normalizing $B\to K^{(*)} \nu\bar{\nu}$ branching ratios by the SM ones and
defining $\mathcal{R}_{K^{(*)}}^\nu \equiv
\mathcal{B}(B\to K^{(*)} \nu\bar{\nu}) / \mathcal{B}(B\to K^{(*)}\nu\bar{\nu})_{\rm SM}$,
we obtain
\begin{align}
 \mathcal{R}_K^\nu = \mathcal{R}_{K^*}^\nu
 = \frac{2}{3} +\frac{1}{3} \left|\frac{ C_L^{\rm SM} +C_L^{\rm NP} }{C_L^{\rm SM}} \right|^2.
\end{align}
%The BABAR collaboration has set the 90\% CL upper limit $\mathcal B(B^+ \to K^+ \bar{\nu}\nu) < 1.7\times 10^{-5}$~\cite{Lees:2013kla}, and 
Combining the charged and neutral $B$ meson decays, the tightest
limits~\cite{Amhis:2016xyh} are set by Belle~\cite{Grygier:2017tzo}, who find
\begin{align}
\mathcal{R}_K^\nu < 3.9, \quad \mathcal{R}_{K^*}^\nu < 2.7
\end{align}
at the 90\% confidence level (C.L.).
The tighter constraint comes from $\mathcal{R}_{K^*}^\nu$, 
and is shown by the cyan lines in Fig.~\ref{const-V-mu}.
The allowed region fully contains the blue hyperbolae favored by 
$b \to s\ell\ell$

%%%%%%%%%%%%%%%%%%%%%%%%%%%%%% ATLAS Z' search
%\subsection{$Z'\to \mu^+\mu^-$ search at LHC}
%\label{sec:pam:Zpmumu}

The $Z'bs$ and $Z'bb$ couplings induce $Z'$ production at LHC via $b\bar s\to Z'$ and
$b \bar b \to Z'$. Hence, with $Z' \to \mu^+\mu^-$, these couplings are constrained 
by dimuon resonance searches at LHC.

We use the results from ATLAS, performed  with 36.1~fb$^{-1}$ 
at 13~TeV~\cite{Aaboud:2017buh} and extract~\cite{extrac}
the 95\% credibility level limit:
$\sigma(pp\to Z'\plusX) \mathcal{B}(Z'\to\mu^+\mu^-) < 42$~fb (9~fb) for $m_{Z'}=200$~GeV (500 GeV).
We calculate the $Z'$ production cross section at leading order (LO) 
using MadGraph5\_aMC@NLO~\cite{Alwall:2014hca} with 
the NN23LO1 parton distribution function (PDF) set~\cite{Ball:2013hta}.
As the ATLAS search does not veto additional activity in the event,
we include also the processes
$gs \to Z'b$, $gb \to Z'b$ and $gg \to Z'b\bar b, Z'b\bar s$ in the cross-section
calculation. We defer the more detailed discussion about  
$Z'$ production at LHC to Section~\ref{coll}.
From the cross sections and the ATLAS limits, we find
\begin{align}
&\sqrt{ |g^L_{bs}|^2 +0.33 |g^L_{bb}|^2} < 0.004 \quad (m_{Z'} = 200~{\rm GeV}), \notag\\
&\sqrt{ |g^L_{bs}|^2 +0.21 |g^L_{bb}|^2} < 0.011 \quad (m_{Z'} = 500~{\rm GeV})
\label{eq:ATLAS-dimuon}
\end{align}
in scenario~(i). 
In scenario~(ii), where the $Z'$ couples to the LH
muon current, the limits are weakened by an overall factor of
$2/\sqrt{3}$ on the right-hand side due to the change in
$\mathcal{B}(Z'\to \mu^+\mu^-)$.

As long as $|g^L_{bb}| \lesssim |g^L_{bs}|$, which is our scenario of interest,
these limits are significantly weaker than those from the $B_s-\bar B_s$ mixing. 
Hence, they are not shown in Fig.~\ref{const-V-mu}. 
For flavor universal models, $Z'$ masses below $m_{Z'}\lesssim 3-4.5$ 
TeV are ruled out by Ref.~\cite{Aaboud:2017buh} with 95\% CL. 
However, a very weakly coupled and flavor non-universal light $Z'$ such as described by  Eq.\eqref{efflag},
escapes the detection and could emerge in the future runs of LHC.

%%%%%%%%%%%%%%%% 
Muon pair production in the scattering of a muon neutrino and a nucleus $N$,
known as neutrino trident production, tightly constrains $Z'\mu\mu$ and 
$Z'\nu_\mu \nu_\mu$ couplings~\cite{Altmannshofer:2014pba}.
The ratio between the total $\nu_\mu N \to \nu_\mu N \mu^+\mu^-$ cross section 
and its SM prediction is given by~\cite{Altmannshofer:2014cfa}
\begin{equation}
\frac{\sigma}{\sigma^{\rm SM}} =
\frac{1 +\left[ 1+4s_W^2 +2(g_{\mu\mu}^V)^2 \frac{v^2}{m_{Z'}^2} \right]^2}{1+(1+4s^2_W)^2},
\end{equation}
in scenario~(i) with $m_{Z'} \gtrsim 10$ GeV.
The measurement~\cite{Mishra:1991bv} by the CCFR collaboration is in 
a good agreement with SM, and implies $\sigma/\sigma^{\rm SM} = 0.82\pm 0.28$.
We show the
resulting $2\sigma$ upper limits on $|g^V_{\mu\mu}|$ by the horizontal
solid red lines in Fig.~\ref{const-V-mu}.

The couplings of the $Z$ boson with the muon and muon neutrino are
modified by $Z'$-loop contributions, which can lead to violation of the
lepton-flavor universality in $Z$ decays. 
In scenario~(i), the vector and axial-vector
$Z\mu\mu$ couplings relative to the SM-like $Zee$ couplings are given
by~\cite{Altmannshofer:2014cfa,Altmannshofer:2016brv}
\begin{align}
\frac{ g_{V\mu} }{ g_{Ve} } 
\simeq \frac{ g_{A\mu} }{ g_{Ae} } 
\simeq 1+ \frac{ (g^V_{\mu\mu})^2 }{16\pi^2}
 {\rm Re}\left[ \mathcal{K}(m_Z^2/m_{Z'}^2) \right], \label{zmumu}
\end{align}
where $\mathcal{K}(m_Z^2 / m_{Z'}^2)$ is a loop function given in Ref.~\cite{Haisch:2011up},
and its real part is taken to match the convention of Ref.~\cite{ALEPH:2005ab}.
Here, the lepton-flavor universality in the SM case is exploited. 
Similarly, normalized $Z\nu\nu$ couplings are given by
\begin{align}
\frac{ g_{V\nu} }{ g_{Ae} } = \frac{ g_{A\nu} }{ g_{Ae} } 
\simeq -\left\{ 1+ \frac{1}{3}\frac{ (g^L_{\mu\mu})^2 }{16\pi^2}
 {\rm Re}\left[ \mathcal{K}(m_Z^2/m_{Z'}^2) \right] \right\},
\label{znunu}
\end{align}
where the factor of $1/3$ effectively takes into account
the fact that only $Z \to \nu_\mu \bar \nu_\mu$ is affected by the $Z'$
among the three neutrino modes. 

The $Z$ couplings were very precisely measured at SLC and LEP.
Relevant results from the average of 14 electroweak measurements 
are $g_{Ve} = -0.03816\pm 0.00047$, $g_{Ae} = -0.50111\pm 0.00035$,
$g_{V\mu} = -0.0367\pm 0.0023$, $g_{A\mu} = -0.50120\pm 0.00054$ and
$g_{V\nu} = g_{A\nu} = 0.5003\pm 0.0012$~\cite{ALEPH:2005ab}.
Of the four possible $Z$ coupling ratios, we take only the one which is most sensitive
to the effect of the $Z'$, i.e. $g_{A\mu}/g_{Ae}= 1.00018 \pm 0.00128$,
where the uncertainties are added in quadrature.
The resulting $2\sigma$ upper limits on $|g^V_{\mu\mu}|$ are shown by
the horizontal red dashed lines in Fig.~\ref{const-V-mu}.

 Nonzero values of $g_{bs}^L$ or $g_{bb}^L$ can alter the $Zbb$
  and $Zss$ couplings at one loop. Taking the $b$ and $s$ quarks to
  be massless, %and  utilizing Eq.~\eqref{zmumu},
  we find that the $Z'$ loop with a nonzero $g_{bs}^L$ modifies the LH $Zbb$ and $Zss$ 
  couplings $g_{Lb}$ and $g_{Ls}$ relative to their SM values $g_{Lb}^{\rm SM}$ and $g_{Ls}^{\rm SM}$  in the same way as
  the LH $Z\mu\mu$ coupling, but with the replacement $g_{\mu\mu}^L \to g_{bs}^L$:
\begin{align}
\frac{ g_{Lb} }{ g_{Lb}^{\rm{SM}} } \simeq \frac{ g_{Ls} }{ g_{Ls}^{\rm{SM}} } \simeq
1+ \frac{ (g^L_{bs})^2 }{16\pi^2}
 {\rm Re}\left[ \mathcal{K}(m_Z^2/m_{Z'}^2)\right]. \label{zbb}
\end{align}
The RH counterparts remain unchanged.
The effect of the $Z'$ loop can be constrained by comparing the measured value
$g_{Lb} = -0.4182\pm0.0015$ ($g_{Ls} = -0.423\pm0.012$)~\cite{ALEPH:2005ab}
to the corresponding SM prediction
$g_{Lb}^{\rm{SM}} = -0.42114^{+0.00045}_{-0.00024}$ 
($g_{Ls}^{\rm{SM}} = -0.42434^{+0.00018}_{-0.00016}$), derived from 
the SM $Z$-pole fit~\cite{ALEPH:2005ab}.
Since $g_{Lb}$ is more precisely measured than $g_{Ls}$,\footnote{ Furthermore,
the measured value of $g_{Ls}$ in Ref.~\cite{ALEPH:2005ab} is obtained 
under the assumption of $g_{Ls} = g_{Ld}$.
This is not valid in our case, since $g_{Ld}$
receives no correction from $Z'$ at one loop.}
we use $g_{Lb}$ to extract the %upper 
limit on $g_{bs}^L$.
Adding the errors in $g_{Lb}$ and $g_{Lb}^{\rm{SM}}$
in quadrature after symmetrizing the $g_{Lb}^{\rm{SM}}$ errors,
we find the $2\sigma$ upper limit %limits 
$|g_{bs}^L| \lesssim 0.34~(0.67)$ for $m_{Z'}=$ 200 (500) GeV.
These limits are much weaker than the ones obtained
from the $B_s$ mixing, and we do not display them in Fig.~\ref{const-V-mu}.
A similar conclusion can be made for the $g_{bb}^L$ coupling as well.

 If both $g_{bs}^L$ and $g_{bb}^L$ are nonzero, an FCNC decay $Z
  \to b \bar s$, which is absent in the SM at tree-level, is induced
  by the one-loop $Z'$ contribution.  A preliminary result by
  DELPHI~\cite{Fuster:1999dj} sets the 90\% CL upper limit $R_{b\ell}
  = \sum_{q = d,s} \sigma(e^+e^- \to b \bar q + \bar b q)/
  \sigma(e^+e^- \to {\rm hadrons}) \leq 2.6\times 10^{-3}$ 
  at the energy scale of the $Z$ mass.  Using
  $\mathcal{B}(Z \to {\rm hadrons})\simeq 70$\%~\cite{Olive:2016xmw},
  one may rewrite the limit as $\sum_{q = d,s} \mathcal{B}(Z \to b
  \bar q + \bar b q) \lesssim 1.8\times 10^{-3}$.  Since the
  $Z'$-loop-induced LH $Zbs$ coupling is suppressed by the factor
  $g_{bs}^L g_{bb}^L/(16\pi^2)$, the DELPHI limit is relevant only if
  both $g_{bs}^L$ and $g_{bb}^L$ are $\mathcal{O}(1)$ and the $Z'$
  mass is not far from the $Z$ mass. Since the $B_s$-mixing constraint
  on $g_{bs}^L$ is much tighter, and we concentrate on the case
  $|g_{bb}^L| \lesssim |g_{bs}^L|$, the impact on $\mathcal{B}(Z \to b \bar s +
  \bar b s)$ is generically far below the DELPHI limit in the scenarios
  considered in this paper.

% %
% \tcr{[MK: If TM is sure about numerics, the following sentence can be 
% incorporated into the previous paragraph by adding the reference info 
% of "Yue:2002ka". 
% (Note this ref. contains the effects of both LH and RH $Z'bs/Z'bb$ as well as
% $Z'ss$ couplings. So, we have to make sure for interpreting their result.)
% If not, we can remove the following sentence.]
% For instance, for $g_{bs}^L =0.001$, $g_{bb}^L = 0.002$ and $m_{Z'} = 200$ GeV,
% we find $\mathcal{B}(Z \to b \bar s + \bar b s) \sim 10^{-16}$,
% using the formula in Ref.~\cite{Yue:2002ka}.}

%%%%%%%%%%% 
The $Z'$ one-loop contribution to the muon anomalous magnetic moment
$a_\mu = (g_\mu - 2)/2$ is~\cite{Pospelov:2008zw}
\begin{align}
 \Delta a_{\mu}= \frac{ (g^V_{\mu\mu})^2}{12 \pi^2}\frac{m_\mu^2}{m_{Z'}^2}, \label{gm2}
\end{align}
where scenario~(i) 
%$g_{\mu\mu}^L = g_{\mu\mu}^R = g_{\mu\mu}^V$ 
and $m_{Z'} \gg m_\mu$ are 
assumed. The difference between the measured value of $a_\mu$ and its SM
prediction is $(2.9\pm0.9)\times 10^{-9}$~\cite{Jegerlehner:2009ry}.
Assigning this difference to Eq.~\eqref{gm2}
yields the dark yellow $2\sigma$ regions in Fig.~\ref{const-V-mu}.
Since the $2\sigma$ constraints from the neutrino trident 
cross section are tighter,
the $Z'$ does not solve the tension in the muon $g - 2$.
At the $3\sigma$ level, the $g_\mu - 2$ regions become compatible
with the neutrino trident production constraints. 
Therefore, we ignore $g_\mu - 2$ in the rest of this paper.

\begin{figure*}[tb!] 
\centering
\includegraphics[width=.360 \textwidth]{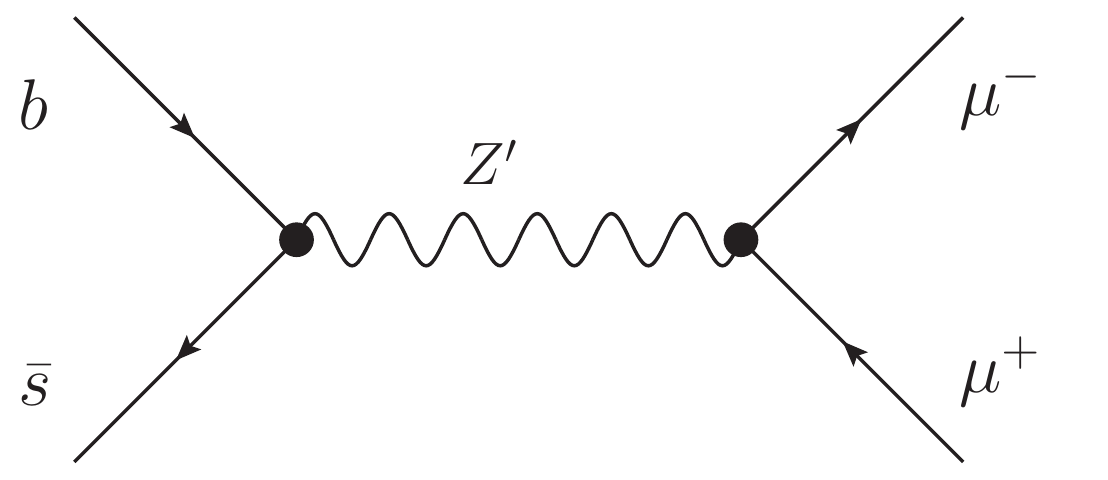}
\caption{A Feynman diagram for the process $b\bar s \to Z' \to \mu^+\mu^-$.}
\label{fig:z}
\end{figure*}
\begin{figure*}[tb!] 
\centering
\includegraphics[width=.580 \textwidth]{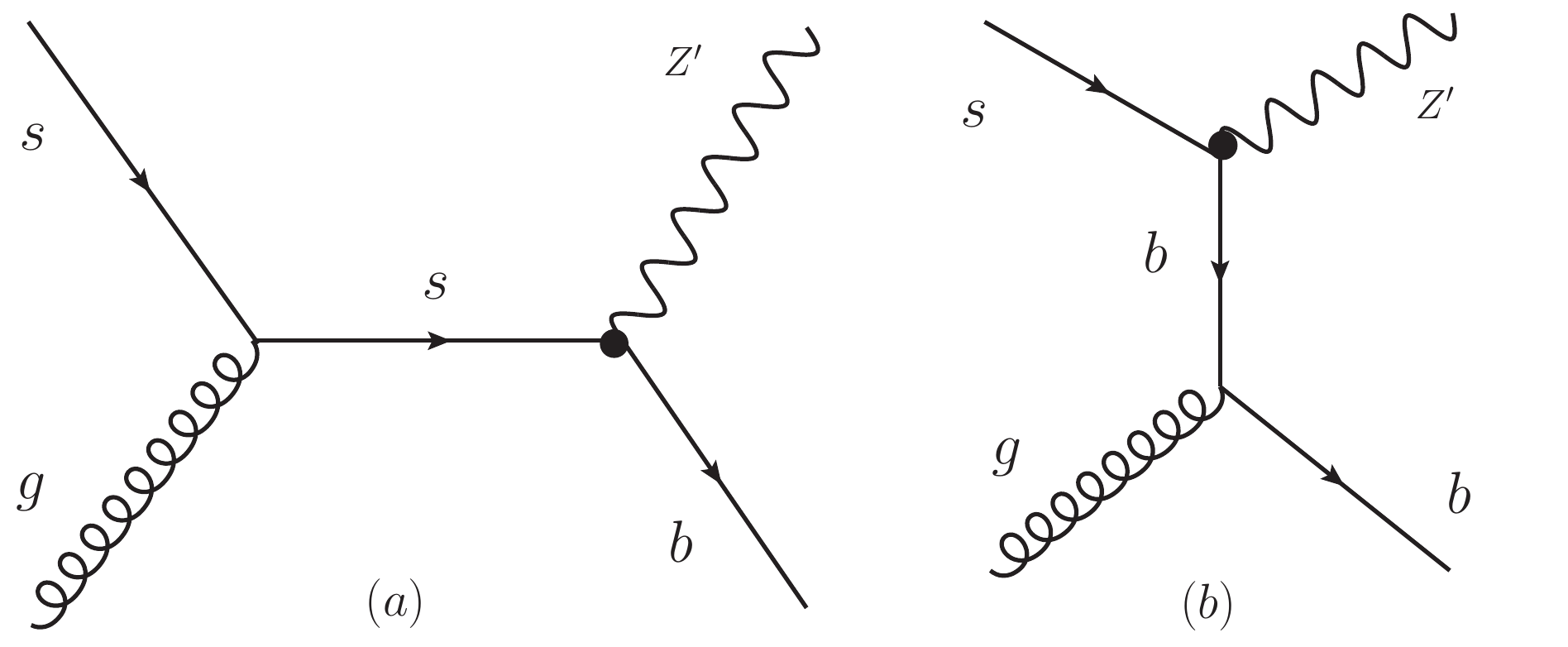}
\caption{Feynman diagrams for $gs \to Z'b$.
} \label{fig:zb}
\end{figure*}

We have discussed so far the constraints in scenario~(i), summarized 
in Fig.~\ref{const-V-mu} on the $g^V_{\mu\mu}$ vs. $g^L_{bs}$ plane. 
From Fig.~\ref{const-V-mu}, we find the constraint on $g^L_{bs}$ is very stringent
due to the $B_s-\bar B_s$ mixing, while the constraint on $g^V_{\mu\mu}$ is much weaker.
With the 30\% NP effects allowed in the $B_s-\bar B_s$ mixing amplitude, 
the former constraint is $|g_{bs}^L| \lesssim 1.4~(3.6) \times 10^{-3}$ for 
$m_{Z'}=200~(500)$ GeV, imposing the lower limit on the $Z'\mu\mu$ coupling
$|g^V_{\mu\mu}| \gtrsim 0.031~(0.078)$ on the $b \to s\ell\ell$ hyperbolae.
This validates the numerical values of the branching ratios in Eq.~(\ref{brs})
to a good approximation, if $|g_{bb}^L|$ is not too large compared to $|g_{bs}^L|$.

The qualitative feature is the same in scenario~(ii), where the $Z'\mu\mu$ coupling
is of LH, as the $B_s-\bar B_s$ mixing constraint does not change. 
Some of the other observables would give slightly different constraints 
on the $Z'\mu\mu$ coupling, but the effect is minor to our study.
The only notable difference from scenario~(i) is the slight change in the value
of $\mathcal{B}(Z'\to \mu^+\mu^-)$ from Eq.~(\ref{brs}).
A nonzero axial-vector $Z'\mu\mu$ coupling can make $B_s \to \mu^+ \mu^-$ relevant,
but the latest LHCb result~\cite{Aaij:2017vad} does not exclude the allowed regions 
for the $b\to s\ell\ell$ anomalies.

%%%%%%%%%%%%%%%%%%%%%%%%%%%%%%%%%%%%%%%%%%%%%%%%%%%%%%%
\section{Discovery and identification of the $Z'$ at LHC
} \label{coll}
%%%%%%%%%%%%%%%%%%%%%%%%%%%%%%%%%%%%%%%%%%%%%%%%%%%%%%%

Having determined the constraints on the $Z'$ couplings, we proceed to study signatures 
for direct production of the on-shell $Z'$ in $pp$ collisions with a center-of-mass energy 
of $\sqrt{s}=14$ TeV.
The goal of this study is to ascertain the LHC potential for both
discovery of the $Z'$ and determination of the flavor structure of its couplings.
Therefore, motivated by the tensions in $b\to s\ell\ell$, we focus on the role of 
the $Z'bs$ coupling $g^L_{bs}$.
If this is the dominant coupling to the quark sector, the $Z'$ will be primarily produced 
via the parton-level process $b\bar{s} \to Z'$ shown in Fig.~\ref{fig:z}.
With the decay $Z' \to \mu^+\mu^-$, the $Z'$ may be discovered in the conventional
dimuon resonance searches.

Such a discovery of a dimuon resonance, however, does not necessarily imply
the existence of the $Z'bs$ coupling. In general, the process
$pp \to Z' \plusX$ may be facilitated by coupling to other quarks,
particularly flavor-diagonal couplings.
To test for dominance of the $Z'bs$ coupling, we propose to also 
search for $pp\to Z'b\plusX$ (see Fig.~\ref{fig:zb} for
typical parton-level processes). 
A $Z'bb$ coupling, predicted in many models (e.g.,
Refs~\cite{Ko:2017quv,Bonilla:2017lsq,Alonso:2017uky,Altmannshofer:2014cfa}),
motivated by the $b \to s\ell\ell$ anomalies, also contribute to
$pp\to Z'\plusX$ and $pp\to Z'b\plusX$ via the parton level processes $b\bar
b\to Z'$ and $gb \to Z' b$.
Since a $Z'bb$ coupling also leads to $pp\to b\bar b Z' \plusX$, via, for example,
$gg \to Z'b\bar b$, measuring the cross section for $pp\to Z'b\bar b \plusX$
may facilitate to discriminate the $Z'bb$ coupling from $Z'bs$.
Similarly, the production process $pp\to Z'b\bar s \plusX$, occurring
due to $gg\to Z'b\bar s$, can in principle help probe the $Z'bs$ coupling.

We explore these signatures using the effective Lagrangian of Eq.~\eqref{efflag}
with scenario~(i), i.e., with a vector $Z'\mu\mu$ coupling.
For each of the two $Z'$ mass values studied, we fix the couplings to the
benchmark points shown by the red dots in Fig.~\ref{const-V-mu}:
\begin{align}
&\left|g^L_{bs}\right|=0.001, \quad \left|g^V_{\mu\mu}\right| = 0.04 \quad
(m_{Z'} = 200~{\rm GeV}), \notag\\
&\left|g^L_{bs}\right|=0.0025, \quad \left|g^V_{\mu\mu}\right| = 0.1 \quad
(m_{Z'} = 500~{\rm GeV}). \label{eq:BP}
\end{align}
These values are selected such that $g^L_{bs}$ leads to a 15\% 
enhancement in the $B_s-\bar B_s$-mixing amplitude $M_{12}$.
The value of $g^V_{\mu\mu}$ is then chosen so as to lie in the range given by
Eqs.~(\ref{eq:WC-fit-i}) and~(\ref{c9-10-Zp}).
We note that one may take a larger $|g^L_{bs}|$ (with a smaller $|g^V_{\mu\mu}|$),
which would enlarge the $pp\to Z'\plusX$ and $pp\to Z'b\plusX$ cross sections by
up to a factor of two, with the $B_s-\bar B_s$-mixing constraint
saturated at $2\sigma$, i.e. a $\sim$30\% enhancement in $M_{12}$.

For the $Z'bb$ coupling, we study three cases for each benchmark
point.  The baseline case is $g^L_{bb}=0$, which restricts assumptions
about the $Z'$ couplings to the minimum needed to explain the $b\to s
\ell\ell$ anomalies.
In addition, we also explore the cases $g^L_{bb}=g^L_{bs}$ and $g^L_{bb}=2g^L_{bs}$,
to study the impact of a nonzero $g^L_{bb}$.
These choices of quark
couplings satisfy the dimuon resonance
search limits in Eq.~(\ref{eq:ATLAS-dimuon}) and maintain the $Z'$
branching ratios in Eq.~\eqref{brs}.

\begin{table*}[bt!] %[hbt!]
\centering
\begin{tabular}{|c|c|c|c|c|c|c|c|c|}
\hline
$m_{Z'}$ (GeV)  
          & \multicolumn{3}{c|}{$\sigma_{\rm signal}$ (fb)}
          & \multicolumn{4}{c|}{$\sigma_{\rm background}$ (fb)}  \\
\cline{2-8}
          & $g^L_{bb}=0$  & $g^L_{bb}=g^L_{bs}$ & $g^L_{bb}=2g^L_{bs}$ 
          & DY %$Z/\gamma^*$  
          & $t\bar{t}$ & $Wt$    & $VV$   \\
\hline
\hline
200       &   1.0         & 1.3                 & 2.2 
          & 170           & 41          & 4.1      & 5.1     \\
\hline
500       &   0.27       & 0.33                & 0.50      
          & 14           & 4.3	        & 0.5	   & 1.0  \\
\hline
\end{tabular}
\caption{Cross sections for the signal process $pp\to Z'\plusX$ with
$Z'\to \mu^+\mu^-$, 
  and the dominant backgrounds after the event selection for the benchmark points 
  defined in Eq.~(\ref{eq:BP}) with the three choices
  for $g_{bb}^L$. The combined cross sections for $WW$, $WZ$ and $ZZ$ backgrounds are denoted together as $VV$. 
}\label{mumu-sigma}
\end{table*}
\begin{table*}[bt!]%[hbt!]
\centering
\begin{tabular}{|c|c|c|c|c|}
\hline
$m_{Z'}$ (GeV)  
          & \multicolumn{3}{c|}{Local significance}  \\
\cline{2-4}
          & $g^L_{bb}=0$  & $g^L_{bb}=g^L_{bs}$ & $g^L_{bb}=2g^L_{bs}$   \\
\hline
\hline
200       &   3.7         & 4.9                 & 8.3     \\
\hline
500       &   3.3         & 4.1                 & 6.5   \\
\hline
\end{tabular}
\caption{Local significance $S_l=N_S/\sqrt{N_B}$ for discovery of the
  process $pp\to Z'\plusX$ with
$Z'\to \mu^+\mu^-$, with an integrated luminosity of $3000~{\rm
    fb}^{-1}$, given the signal and background cross
  sections shown in Table~\ref{mumu-sigma}.
}\label{mumu-signif}
\end{table*}

In the following subsections, we mainly focus on the discovery potential of the $Z'$ 
in the production processes $pp \to Z' \plusX$, $pp \to Z'b \plusX$, and $pp \to Z'b\bar{b} \plusX$,
with the $Z'$ always decaying to $\mu^+\mu^-$.
We use Monte Carlo event generator
MadGraph5\_aMC@NLO~\cite{Alwall:2014hca} to generate signal and
background samples at LO with the NN23LO1 PDF set~\cite{Ball:2013hta}. 
The effective Lagrangian of Eq.~\eqref{efflag} is implemented in the
FeynRules~2.0~\cite{Alloul:2013bka} framework. The matrix elements for
signal and background are generated with up to two additional jets and
interfaced with PYTHIA~6.4~\cite{Sjostrand:2006za} for parton
showering and hadronization. Matching is performed with the MLM
prescription~\cite{Alwall:2007fs}. The generated events are passed
into the Delphes~3.3.3~\cite{deFavereau:2013fsa} fast detector
simulation to incorporate detector effects based on ATLAS.

%%%%%%%%%%%%%%%%%%%%%%%%%%%%%%%%%%%%%%%%%%%%%%%%%%%%
\subsection{Observation of $pp\to Z'\plusX$}
\label{subsec:dimuon}
%%%%%%%%%%%%%%%%%%%%%%%%%%%%%%%%%%%%%%%%%%%%%%%5
Several SM processes constitute background for $pp\to Z'\plusX$,
where we remind the reader that the $Z'$ decays into $\mu^+\mu^-$.
The dominant background is due to the Drell-Yan (DY) events,
$pp \to Z/\gamma^*\plusX$. The 
$pp\to t\bar{t}$ events with semileptonic
decay of both top quarks is the next largest background.
Smaller backgrounds arise from $pp\to Wt$ and $VV$, where $V\equiv W, Z$. Background may also arise from
leptons produced in heavy-flavor decays or from jets faking leptons.
These background sources are not well modeled by the simulation tools,
and we ignore them, assuming that they can be reduced to subdominant level 
with lepton quality cuts without drastically impacting the results of our analysis.

We scale the LO cross sections obtained by MadGraph5\_aMC@NLO as follows.
The DY cross section is normalized to a NNLO QCD+NLO EW cross section
by a factor of $1.27$, obtained with FEWZ~3.1~\cite{Li:2012wna} in the
dimuon-invariant mass range $m_{\mu\mu} > 106$~GeV.
We normalize the LO $pp\to t\bar{t}$ and $pp\to Wt$ cross sections to NNLO+NNLL 
cross sections
by $1.84$~\cite{twiki} and $1.35$~\cite{Kidonakis:2010ux}, respectively.
The $pp\to WW$, $pp\to WZ$ and $pp\to ZZ$ cross sections are normalized to NNLO QCD 
by $1.98$~\cite{Gehrmann:2014fva}, $2.07$~\cite{Grazzini:2016swo} 
and $1.74$~\cite{Cascioli:2014yka}, respectively. We do not apply correction factors
to the signal cross sections throughout this paper.

\label{subsec:bmumu}

\begin{table*}[bt!]%[hbt!]
\centering
\begin{tabular}{|c|c|c|c|c|c|c|c|c|c|c|}
\hline
$m_{Z'}$ (GeV)  
          & \multicolumn{3}{c|}{$\sigma_{\rm signal}$ (fb)}
          & \multicolumn{6}{c|}{$\sigma_{\rm background}$ (fb)}  \\
\cline{2-10}
          & $g^L_{bb}=0$  & $g^L_{bb}=g^L_{bs}$ & $g^L_{bb}=2g^L_{bs}$ 
          & ${\rm DY}+b$  & ${\rm DY}+c$ & 
          ${\rm DY}+j$ & $t\bar t$ & $Wt$ & $VV$   \\
\hline
\hline
200       &   0.17        & 0.22                & 0.37 
          & 1.3   & 1.0  & 0.22 & 5.6 & 0.8 & 0.5     \\
\hline
500       &   0.043       & 0.049               & 0.10
          & 0.15  & 0.048 & 0.028 & 0.26 & 0.08 & 0.064    \\
\hline
\end{tabular}
\caption{
Cross sections for the signal process $pp\to Z'b\plusX$ with
$Z'\to \mu^+\mu^-$,
  and the dominant backgrounds after the event selection for the benchmark points 
  defined in Eq.~(\ref{eq:BP}) with the three choices
  for $g_{bb}^L$.
}\label{mumub-sigma}
\end{table*}

We select events that contain at least two oppositely charged muons.
The transverse momentum of each muon is required to satisfy
$p^T_\mu>50$~GeV, and its pseudorapidity must be in the range $|\eta_\mu|<2.5$.
The two muons must satisfy 
$\Delta R_{\mu\mu} = \sqrt{\Delta \phi^2_{\mu\mu} + \Delta \eta^2_{\mu\mu}}> 0.4$, 
where $\Delta \phi_{\mu\mu}$ and $\Delta \eta_{\mu\mu}$ are the
separations in azimuthal angle and pseudorapidity between the muons.
Finally, we require the invariant mass of the two muons to satisfy
$|m_{\mu\mu}- m_{Z'}| < m_{\rm cut}$,
where $m_{\rm cut} =4$~GeV and 16~GeV for $m_{Z'}=200$~GeV and 
$m_{Z'}=500$~GeV, respectively. These values are 
chosen so as to maximize the naive local significance 
of the no-signal hypothesis, $S_l=N_S/\sqrt{N_B}$, where
$N_S$ and $N_B$ are the expected signal and background yields.

The invariant mass cut $|m_{\mu\mu}- m_{Z'}| < m_{\rm cut}$ is not
realistic for a discovery scenario, in which one does not know
the true mass $m_{Z'}$. However, the value of $S_l$ thus obtained is a rough
estimate of the one that will be found by the more sophisticated
analysis that will eventually be performed with the full LHC data.
One is actually interested in the global significance $S_g$, which
accounts for the probability to obtain the given value of $S_l$
anywhere in the dimuon-invariant mass range.  Rigorous methods for
estimating $S_g$ exist~\cite{Cowan:2010js}.
However, at this level of approximation, it is sufficient to use the
crude estimate
\begin{equation}
P_g = P_l {m_{\rm cut} \over m_{\rm range}},
\label{eq:P_g}
\end{equation}
where $P_g$ and $P_l$ are the $\chi^2$ probabilities corresponding to
$S_g$ and $S_l$, respectively, and $m_{\rm range}$ is the size of the
range of $m_{\mu\mu}$ values explored in the analysis. Since cross
sections drop to negligible levels at high $m_{\mu\mu}$, it is
reasonable to take $m_{\mu\mu}\sim 2$--$3$ TeV for this estimate.

The cross sections for signal and backgrounds after the cuts are
listed in Table~\ref{mumu-sigma} for the benchmark points defined in
Eq.~(\ref{eq:BP}) with the three choices of $g_{bb}^L$.  The
corresponding values of the local significance $S_l$ for an integrated
luminosity $L=3000~{\rm fb}^{-1}$ are summarized in
Table~\ref{mumu-signif}.
Inserting the values of $S_l$ into Eq.~(\ref{eq:P_g}), we conclude
that the global significance will likely be greater than $5\sigma$ 
for the case $g^L_{bb}=2g^L_{bs}$, allowing separate discovery by
ATLAS and CMS. For $|g^L_{bb}| \le
|g^L_{bs}|$, the global significance will be under $5\sigma$.
Whether the $5\sigma$ mark will be passed by combining ATLAS
and CMS results is beyond the precision of our rough estimate.

A larger $|g_{bs}^L|$ can enhance the significance in each benchmark scenario.
For the scenario of $m_{Z'}=200~(500)$ GeV with $g_{bb}^L = 0$, 
taking $|g_{bs}^L| = 0.0013~(0.0034)$ with $|g_{\mu\mu}^V|=0.031~(0.074)$ 
pushes the local significance slightly above $6\sigma$,
which may imply a global significance of $5\sigma$.
In this case, the $B_s-\bar B_s$ mixing amplitude $|M_{12}|$ is also enhanced
from the SM one by 25\% (28\%), but still within the nominal 2$\sigma$ allowed range,
as discussed in Sec.~\ref{sec:pam}.

%%%%%%%%%%%%%%%%%%%%%%%%%%%%%%%%%%%%%%%%%%%%%%%%%%%%%%%%%%%%%%55
\subsection{Observation of  $pp\to Z' b\plusX$}
\begin{table*}[bt!]%[hbt!]
\centering
\begin{tabular}{|c|c|c|c|c|}
\hline
$m_{Z'}$ (GeV)  
          & \multicolumn{3}{c|}{Local significance}  \\
\cline{2-4}
          & $g^L_{bb}=0$  & $g^L_{bb}=g^L_{bs}$ & $g^L_{bb}=2g^L_{bs}$   \\
\hline
\hline
200       &   3.0         & 3.9                 & 6.6     \\
\hline
500       &   3.0         & 3.4                 & 7.2   \\
\hline
\end{tabular}
\caption{
Local significance $S_l=N_S/\sqrt{N_B}$ for discovery of the
  process $pp\to Z'b\plusX$ with
$Z'\to \mu^+\mu^-$, with an integrated luminosity of $3000~{\rm
    fb}^{-1}$, given the signal and background cross
  sections shown in Table~\ref{mumub-sigma}.
}\label{mumub-signif}
\end{table*}

The main SM background for $pp\to Z' b\plusX$ is $pp\to t\bar{t}$. The
second-largest background is Drell-Yan with at least one additional $b$ jet,
labeled as ${\rm DY}+b$.  Smaller contributions arise from ${\rm DY}+c$,
$pp\to Wt$, $pp\to VV$, and ${\rm DY}+j$, where $j$ stands for a jet
from a gluon or a $u$, $d$, or $s$ quark.
We normalize the cross sections for ${\rm DY}+b$, ${\rm
  DY}+c$, and ${\rm DY}+j$ to 
NNLO QCD by $1.83$~\cite{Boughezal:2016isb}. The
correction factors for $pp\to t\bar{t}$, $pp\to Wt$, and $pp\to VV$ are taken to be
the same as in Section~\ref{subsec:dimuon}.

We select simulated events that contain at least two opposite-charge
muons.  The muons are required to be in the pseudorapidity range
$|\eta_\mu| < 2.5$, have minimal transverse momenta of $p^T_\mu >
50~(60)$ GeV for $m_{Z'}=200~(500)$ GeV, and be separated by $\Delta
R_{\mu\mu} > 0.4$. Jets are reconstructed using the anti-$k_T$
algorithm with radius parameter $R=0.5$.  
It is assumed that a $b$-tagging algorithm reduces the efficiency
for $c$~jets and light jets by factors of 5 and 137,
respectively~\cite{ATLAS:2014ffa}. Its efficiency for $b$~jets is
calculated in Delphes, accounting for the $p_T$ and $\eta$ dependence.
The leading $b$ jet is required to have transverse momentum $p^T_b>30$~GeV with
$|\eta_b| < 2.5$, and its separation from each of the two leading
muons must satisfy $\Delta R_{b\mu} > 0.4$. We reject events that have
a second $b$-tagged jet with $p^T_b>30$~GeV, slightly increasing
the local significance.  The missing transverse energy must be less
than $40$~GeV, in order to reduce $pp\to t\bar{t}$ and $pp\to Wt$ backgrounds.  
Finally,
we apply the optimized dimuon-invariant mass cut $|m_{\mu\mu}- m_{Z'}|
< 5~(15)$~GeV for $m_{Z'}=200~(500)$~GeV.

The resulting cross sections are shown in Table~\ref{mumub-sigma}, and 
the corresponding local signal significances with 3000 fb$^{-1}$
are summarized in Table~\ref{mumub-signif}.
The local significances are slightly smaller than the corresponding ones 
in Table~\ref{mumu-signif}, 
except in the case of $m_{Z'}=500$ GeV and $g_{bb}=2g_{bs}^L$.
Thus, we conclude that, like $pp\to Z' \plusX$, the process
$pp\to Z'b \plusX$ is likely be discovered at $S_g>5$ if $g_{bb}^L\ge 2 g_{bs}^L$
in our benchmark points. By scaling the values in Table~\ref{mumub-signif}, 
we observe that a local significance of $6\sigma$
can be attained with $|g_{bs}^L|\gtrsim 0.0014~(0.0036)$
for $m_{Z'}=200~(500)$~GeV, even if $g_{bb}^L=0$, 
at the cost of a $\sim 30$\% enhancement in $|M_{12}/M_{12}^{\rm SM}|$.

%%%%%%%%%%%%%%%%%%%%%%%%%%%%%%%%%%%%%%%%%%%%%%%%%%%%%%%%%%%%%%
\subsection{Observation of $pp\to Z' b \bar b \plusX$ and $pp\to Z' b \bar s \plusX$}
\label{subsec:bbmumu}

\begin{table*}[bt!]%[hbt!]
\centering
\begin{tabular}{|c|c|c|c|c|c|c|c|c|c|c|}
\hline
$m_{Z'}$ (GeV)  
          & \multicolumn{3}{c|}{$\sigma_{\rm signal}$ (fb)}
          & \multicolumn{3}{c|}{$\sigma_{\rm background}$ (fb)}  \\
\cline{2-7}
          & $g^L_{bb}=0$  & $g^L_{bb}=g^L_{bs}$ & $g^L_{bb}=2g^L_{bs}$ 
          & ${\rm DY}+\mbox{h.f.}$ jets  & $t\bar t$ & $Wt$    \\
\hline
\hline
200       &  0.00018        &  0.0025       & 0.0094
          & 0.2  & 1.5  &  0.5     \\
\hline
500       &   0.00008   &  0.0006 & 0.0026
          & 0.07  & 0.27 &  0.17     \\
\hline
\end{tabular}
\caption{
Cross sections for the signal process $pp\to Z'b\bar{b}\plusX$ with
$Z'\to \mu^+\mu^-$,
  and the dominant backgrounds after the event selection for the benchmark points 
  defined in Eq.~(\ref{eq:BP}) with the three choices
  for $g_{bb}^L$.
}\label{mumubb-sigma}
\end{table*}
\begin{table*}[htpb!]%[hbt!]
\centering
\begin{tabular}{|c|c|c|c|c|}
\hline
$m_{Z'}$ (GeV)  
          & \multicolumn{3}{c|}{Local significance}  \\
\cline{2-4}
          & $g^L_{bb}=0$  & $g^L_{bb}=g^L_{bs}$ & $g^L_{bb}=2g^L_{bs}$   \\
\hline
\hline
200       & 0.007          &  0.1               &  0.35    \\
\hline
500       & 0.006          &  0.05              &  0.2 \\
\hline
\end{tabular}
\caption{
Local significance $S_l=N_S/\sqrt{N_B}$ for discovery of the
  process $pp\to Z'b\bar b\plusX$ with
$Z'\to \mu^+\mu^-$, with an integrated luminosity of $3000~{\rm
    fb}^{-1}$, given the signal and background cross
  sections shown in Table~\ref{mumubb-sigma}.
}\label{mumubb-signif}
\end{table*}

The dominant SM backgrounds for $pp\to Z' b \bar b \plusX$
are $pp\to t \bar t$, $pp\to Wt$ and ${\rm DY}+b$ or $c$ jets.
$pp\to VV$ gives a negligible contribution.
We adopt the same correction factors for the background cross sections and
follow the same event selection criteria as in Section~\ref{subsec:bbmumu},
and in addition require 
the subleading $b$ jet to have $p^T_b > 30$~GeV, $|\eta_{b}| < 2.5$,
and to be separated from the leading $b$~jet and each of the two leading muons by
$\Delta R > 0.4$, for both the $Z'$ masses.

The resulting cross sections are shown in Table~\ref{mumubb-sigma}.
The $g^L_{bb}= 0$ cases have tiny but nonzero cross sections, due to
production via $b \bar s \to Z'g^*$, with $g^*\to b \bar b$.  By
contrast, a nonzero $g^L_{bb}$ induces the less suppressed process
$gg\to Z' b \bar b$.  The corresponding local signal significances for
3000~fb$^{-1}$ are given in Table~\ref{mumubb-signif}.
As the local significances are much less than 1, we conclude that 
observation of this process is not possible at LHC
within the range of couplings explored here.

In general, $|g_{bb}^L|$ may take a larger value, up to the limit of
Eq.~(\ref{eq:ATLAS-dimuon}), namely, $|g^L_{bb}| = 0.007$ ($0.024$)
for $m_{Z'}= 200$ (500)~GeV with $g^L_{bs} \sim 0$. With these values,
we estimate the cross section of $pp\to Z' b \bar b \plusX $ to be
0.098~fb (0.055~fb) after the event selection cuts.  This corresponds
to a local significance of around $3.6\sigma$ ($4.1\sigma$) for an
integrated luminosity of $3000~{\rm fb}^{-1}$. Thus, a global
significance of $5\sigma$ is not likely. 
We note, however, that since we used the same QCD correction factors for the
background cross sections as in the $pp\to Z'b \plusX$ case, there
is a greater uncertainty on these cross sections.

Generally, the cross section for $pp\to Z' b \bar b \plusX$
is strongly suppressed by the 3-body phase space. 
Since the same suppression applies for  $pp\to Z' b \bar s \plusX$,
one expects the cross section for this process 
to be small as well. Moreover, the process $pp\to Z' b \bar s \plusX$
would also suffer from light-jet backgrounds which make the discovery 
not possible, given the $B_s-\bar B_s$ mixing constraint.

%%%%%%%%%%%%%%%%%%%%%%%%%%%%%%%%%%%%%%%%%%%%%%%%%%%%%
%%%%%%%%%%%%%%%%%%%%%%%%%%%%%%%%%%%%%%%%%%%%55
\section{Impact of the right-handed $Z'bs$ coupling}
\label{sec:RHb2s}

\begin{figure*}[t!]
\centering
 \includegraphics[width=.46 \textwidth]{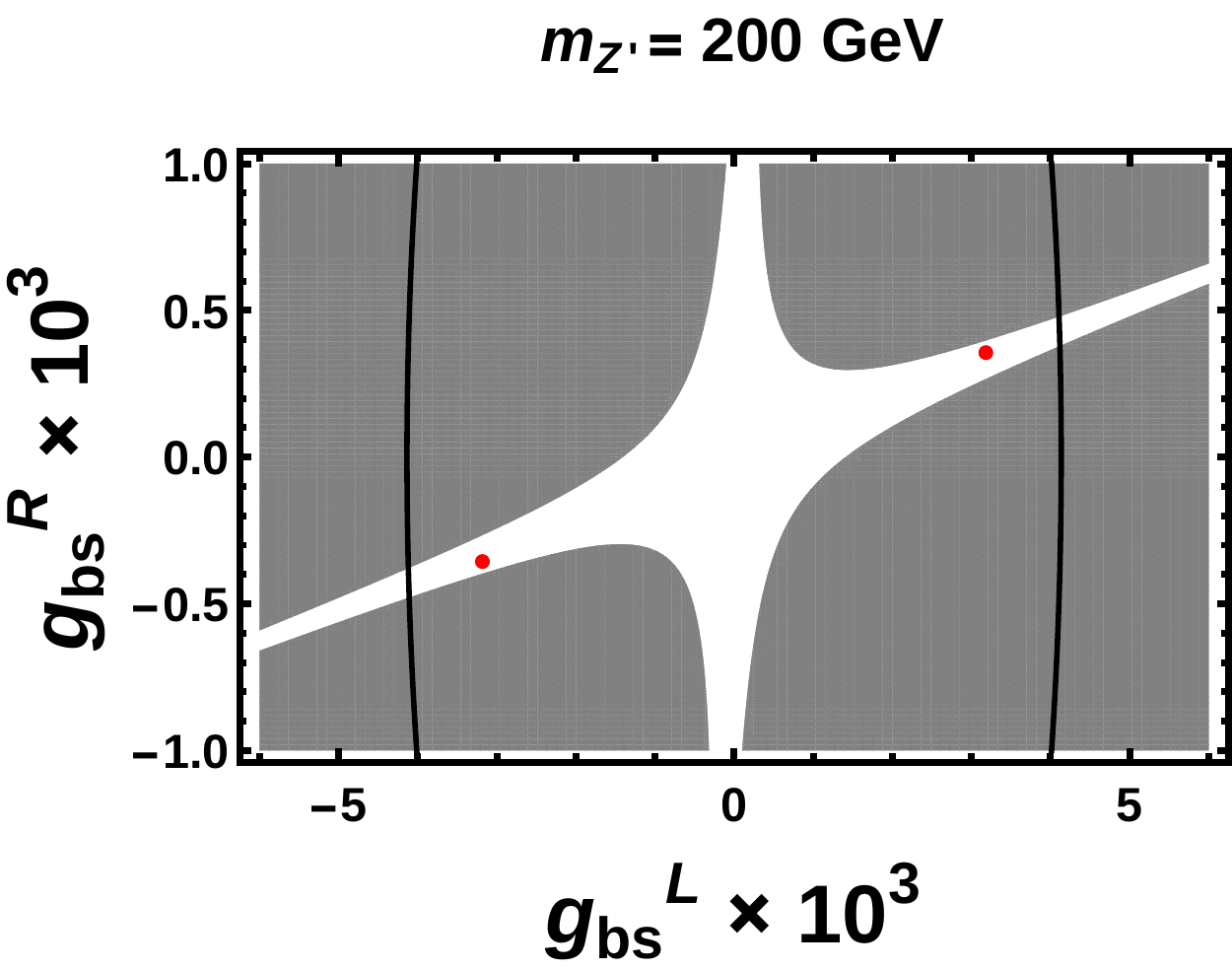}
 \includegraphics[width=.44 \textwidth]{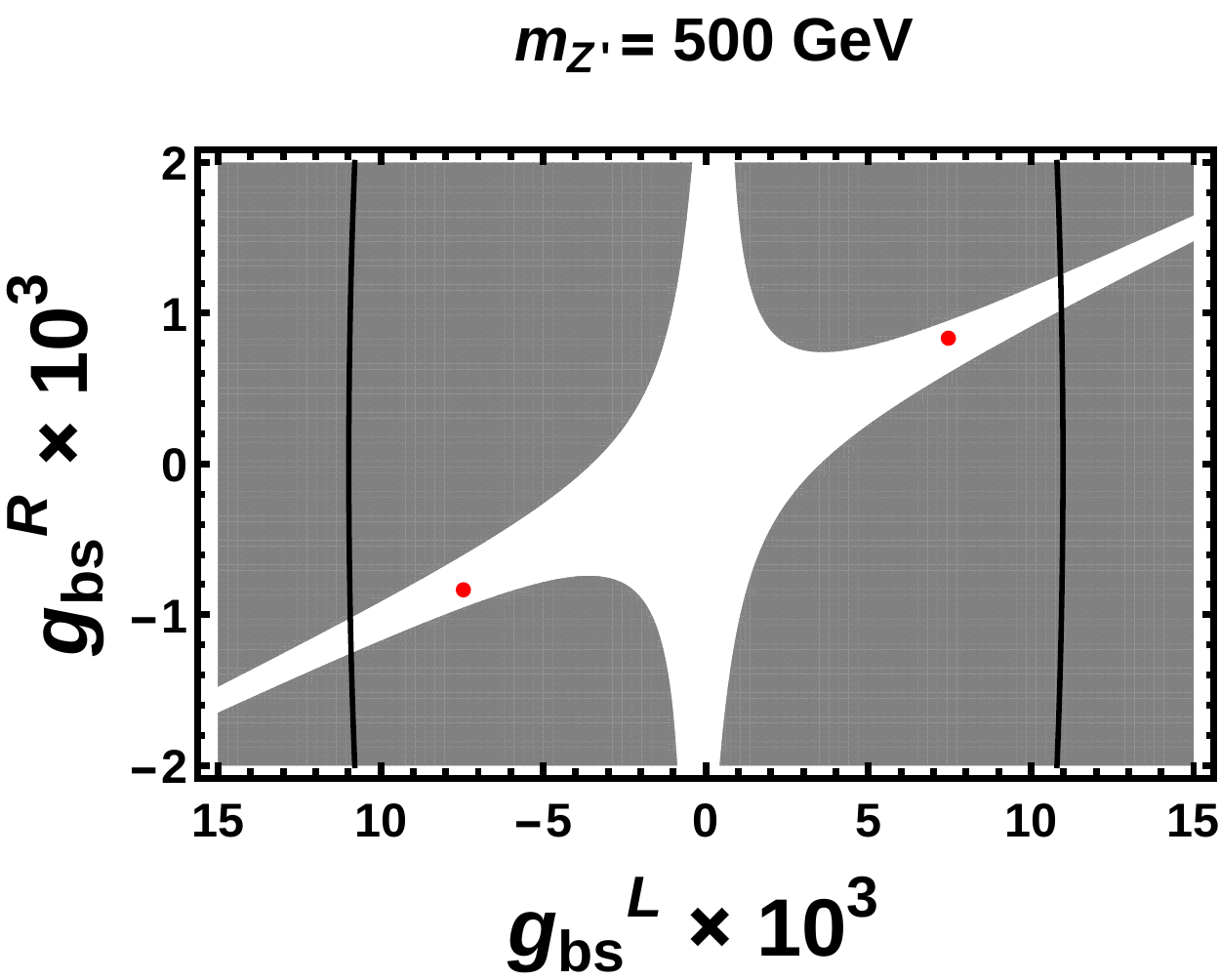}
\caption{ 
Constraints on $g^R_{sb}$ vs. $g^L_{sb}$ (in units of $10^{-3}$) for
$m_{Z'} = 200$ GeV [left] and 500 GeV [right], when allowing the
$B_s-\bar B_s$-mixing amplitude
to deviate by up to 30\% from its SM
prediction.  The gray regions are excluded.  The solid lines show the
exclusion by the ATLAS dimuon resonance search~\cite{Aaboud:2017buh}
for $\mathcal{B}(Z' \to \mu^+\mu^-) \simeq 2/3$.  The red dots show our
benchmark points.
}\label{allowed_gbslr}
\end{figure*}

In this section, we study an impact of a tiny but nonzero RH $Z'bs$ coupling
by adding the following terms to the effective Lagrangian in Eq.~\eqref{efflag}:
\begin{align}
\Delta \mathcal{L} = 
 -g^R_{bs} \left( \bar{b}\gamma^\alpha P_R s + \bar{s}\gamma^\alpha P_R b \right)Z'_\alpha.
\end{align}
The resulting additional contributions to $b \to s \mu^+\mu^-$ are
described by effective operators as in Eq.~(\ref{eq:bsll}), with $P_L$
replaced by $P_R$, and with $C_9^{\rm NP}$ and $C_{10}^{\rm NP}$ replaced by the
Wilson coefficients
\begin{align}
C'_9 =  \frac{g^{R}_{bs}~g^{V}_{\mu\mu}}{\mathcal{N} m_{Z'}^2}, \quad
C'_{10} =  \frac{g^{R}_{bs}~g^{A}_{\mu\mu}}{\mathcal{N} m_{Z'}^2}. \label{c910pm}
\end{align}
There is no significant indication for nonzero $C'_{9,10}$ in
the majority of the $b\to s\ell\ell$ global fit analyses. However, even a
tiny $g^{R}_{bs}$ can drastically affect the $B_s-\bar B_s$ mixing,
which is now given by~\cite{Altmannshofer:2014cfa}
\begin{align}
 \frac{M_{12}}{M_{12}^{\rm SM}}=1+\frac{1}{m_{Z'}^2}
 \left[ (g^L_{bs})^2 -9.7 g^L_{bs}g^R_{bs} +(g^R_{bs})^2 \right]\nn\\
 \times\left(\frac{g^2_2}{16\pi^2 v^2}(V_{tb}V_{ts}^*)^2 S_0\right)^{-1}, \label{eq:M12-RH}
\end{align}
calculated with the hadronic matrix elements in Ref.~\cite{Buras:2012jb}.
The large negative coefficient of the $g^L_{bs}g^R_{bs}$
term, which is partly due to
renormalization group effects~\cite{Buras:2001ra},
means that a small value of $g^R_{bs}$ allows for a large $g^L_{bs}$,
due to cancellation between the terms.
In Fig.~\ref{allowed_gbslr} we show the $B_s-\bar B_s$-mixing
constraint on $g^R_{bs}$ vs. $g^L_{bs}$, when $|M_{12}|$ is allowed to
change by up to 30\% of its SM value.
Reducing the allowed NP contribution to $|M_{12}|$, say, to 15\%,
would narrow the width of the
tilted-cross-shaped allowed region in Fig.~\ref{allowed_gbslr}. However, it
would not change the conclusion, namely, that a large value of $|g^L_{bs}|$ is
allowed.
What now becomes the most significant limit on $g^L_{bs}$ is the ATLAS
dimuon resonance search~\cite{Aaboud:2017buh}, shown by the solid
lines, assuming Eq.~(\ref{brs}).

The cancellation in $M_{12}$ requires $g_{bs}^L g_{bs}^R > 0$. This implies
$({\rm Re}~C_9^{\rm NP})({\rm Re}~C'_9) > 0$,
contrary to the best-fit values for $C_{9}^{\rm NP}$ and $C'_9$, 
e.g., Refs.~\cite{Capdevila:2017bsm,Altmannshofer:2017yso}.
However, the cancellation requires only $g_{bs}^R \sim 0.1 g_{bs}^L$
or $C'_9 \sim 0.1 C_9^{\rm NP}$. While the fits favor $C_9^{\rm NP} \sim -1$
and are consistent with $C'_9 = 0$, they cannot exclude
a small  negative $C_9'$. 
Indeed, the point $(C_9^{\rm NP}, C'_9) = (-1.11, -0.1)$ is at the border of
the 1$\sigma$ ellipse in Ref.~\cite{Capdevila:2017bsm}
with the assumption $C_{10}^{\rm NP}=C'_{10}=0$.

To illustrate the impact of such a possibly large $Z'bs$ coupling on the
$Z'$ discovery potential, we consider scenario~(i) with the following
benchmark points for $m_{Z'}= 200$ and $500$ GeV respectively (corresponding to the dots in
Fig.~\ref{allowed_gbslr}):
\begin{align}
&\left|g^L_{bs}\right| = 0.0032, \left|g^R_{bs}\right| = 0.00036,  
\left|g^V_{\mu\mu}\right| = 0.012, \notag\\
&\left|g^L_{bs}\right|=0.0075, \left|g^R_{bs}\right|=0.00083, 
 \left|g^V_{\mu\mu}\right| = 0.032. \label{eq:BP-RH}
\end{align}
Both points correspond to $(C_9^{\rm NP}, C'_9) = (-1.11, -0.1)$.
The effects of the tiny $g^R_{bs}$ on other constraints, 
e.g. $B\to K^{(*)}\nu\bar\nu$, are negligible.
Taking $g_{bb}^L = 0$, we find that 
the $pp\to Z'\plusX$ and $pp\to Z' b\plusX$ cross sections are highly enhanced 
compared to the ones in Sec.~\ref{coll},
while the relatively large $g^L_{bs}$ values lead to a non-negligible $Z' \to b\bar s$
branching ratio, slightly reducing $\mathcal{B}(Z'\to \mu^+\mu^-)$ from $\sim66\%$ to $\sim52\%$ and $\sim55\%$
respectively for 200 and 500 GeV $Z'$.

\begin{table}[h!]
\centering
\begin{tabular}{|c|c|c|c|}
\hline
$m_{Z'}$ (GeV)  
          & \multicolumn{2}{c|}{Local significance}  \\
\cline{2-3}
          & $\mu^+ \mu^- \plusX$  & $\mu^+\mu^- b \plusX$   \\
\hline
\hline
200 & 9.3 & 7.6 \\
\hline
500 & 7.7 & 6.9 \\
\hline
\end{tabular}
\caption{Local significance $N_S/\sqrt{N_B}$ for discovery of the process 
$pp\to Z'\plusX\to \mu^+\mu^- \plusX$ and $pp\to Z'b\plusX \to \mu^+\mu^-b\plusX$
with the integrated luminosity of $300~{\rm fb}^{-1}$, obtained by rescaling 
the results of Table \ref{mumu-signif} and \ref{mumub-signif} 
to the benchmark points defined in Eq.~(\ref{eq:BP-RH}) for $g_{bb}^L=0$.
}\label{RH-signif}
\end{table}

Taking account of these effects and rescaling the significances obtained in Sec.~\ref{coll},
we show in Table~\ref{RH-signif} the local significances for discovery of 
$pp\to Z'\plusX$ and $pp\to Z'b\plusX $ with the
integrated luminosity of $300~{\rm fb}^{-1}$, for the two benchmark points defined above.
The results suggest that both processes can be discovered with the Run-3 dataset,
even if the $Z'bb$ coupling vanishes. A larger 
$\left|g^L_{bs}\right|$ in this scenario enhances the cross section for $pp\to Z'b\bar s\plusX$ process, but
the discovery would still be beyond the reach of the HL-LHC. 

%%%%%%%%%%%%%%%%%%%%%%%%%%%%%%%%%%%%%%%%%%%%%%%%%%%%%%%%%%%%%%%%%%%%%%%%%%%%
\section{Summary and Discussions}
\label{summ}

Observed tensions in $b\to s\ell\ell$ measurements can be explained by 
a new $Z'$ boson that couples to the left-handed $b \to s$ current as well as to muons.
In this paper, we have studied the collider phenomenology of such a $Z'$ 
based on an effective model introduced in Eq.~(\ref{efflag}).
For this purpose, we first estimated phenomenological constraints on 
the $Z'bs$ and $Z'\mu\mu$ couplings 
for the representative masses $m_{Z'}=200$ and 500 GeV. 
The most important constraint is the $B_s$ mixing, 
which tightly constrains the LH $Z'bs$ coupling $g^L_{bs}$.
For fixed values of $g^L_{bs}$ and $m_{Z'}$, the allowed coupling to muons 
is determined by global fits to $b\to s\ell\ell$ data,
up to the value allowed by the constraint from the neutrino trident production, where 
the $Z'\nu_\mu\nu_\mu$ coupling is related to $Z'\mu\mu$ coupling by the SU(2)$_L$ symmetry.
We also introduced the $Z'bb$ coupling $g^L_{bb}$, which is mildly constrained 
by the dimuon resonance search at LHC.
The resulting couplings are such that the $Z'$ decays mostly to $\mu^+\mu^-$ 
and $\nu_\mu\bar\nu_\mu$, with the two branching ratio values mildly depending
on whether the muon coupling is vector-like or left-handed.

Given the coupling constraints, we explored the capability of the 14 TeV LHC
to discover the $Z'$ and to determine the flavor structure of its couplings.
For the sake of this dual goal, we studied the two processes 
$pp\to Z'+X\to \mu^+\mu^-+X$ and $pp\to Z'b+X \to  \mu^+\mu^- b +X$,
where the former may be induced by $b\bar{s} \to Z'$ and/or $b\bar{b} \to Z'$
and the latter by $gs \to Z'b$ and/or $gb \to Z'b$. We considered
two representative  $Z'$ masses of $200$ and $500$ GeV with three scenarios for the
$Z'bb$ coupling: $g_{bb}^L = 0$, $g_{bs}^L$ or $2g_{bs}^L$.
For $g_{bb}^L = 0$, we found that discovery of $pp\to Z'+X$ and $pp\to Z'b+X$  (with about 5$\sigma$ global significance) 
with 3000 fb$^{-1}$ data requires a large $Z'bs$ coupling, so that the $B_s$ mixing amplitude $M_{12}$ is enhanced by 
$\sim 30$\% or more relative to the SM expectation. This corresponds roughly to the
2$\sigma$ upper limits of the global analyses~\cite{Charles:2015gya,Bona:2006sa} for the CKM parameters.
With a nonzero $g^L_{bb}$, discovery in both the modes is possible without 
such a drastic effect on the $B_s$ mixing; in particular, for $g_{bb}^L = 2g_{bs}^L$, 
discovery is possible with a $\sim 15$\% deviation in the $M_{12}$.
For further discrimination between the $Z'bs$ and $Z'bb$ couplings,
we also studied the process $pp\to Z'b \bar b +X \to \mu^+\mu^- b \bar b+X$,
predominantly arising from $Z'bb$ coupling.
However, we found it to be not promising even with 3000 fb$^{-1}$ integrated luminosity, due primarily to three-body phase-space suppression.
The same conclusion applies to $pp\to Z'b \bar s \plusX$, which gives direct access to the $Z'bs$ coupling.

The discovery potential of the $Z'$ is rather limited due to the $B_s$ mixing constraint.
The $B_s$ mixing constraint, however, is only indirect and is susceptible to
the details of the UV completion of the effective model.
In particular, we illustrated that the existence of a tiny but nonzero right-handed $Z'bs$ coupling 
$g_{bs}^R$ accommodates a large LH $Z'bs$ coupling due to the cancellation 
in the $B_s$ mixing amplitude, without conflicting with the $b \to s \ell\ell$ global fits.
In this case we found that discovery in both the $pp\to Z'+X$ and $pp\to Z' b+X$ processes
may occur even with $\mathcal{O}(100)$ fb$^{-1}$ integrated luminosity.

Comments on the subtlety of the implementation of the $B_s$ mixing constraint
are in order (see also Sec.~\ref{sec:pam}). 
As mentioned above, a 30\% enhancement in the $B_s$ mixing amplitude $M_{12}$ by NP
roughly corresponds to the 2$\sigma$ upper limits by the latest global analyses 
of CKMfitter~\cite{Charles:2015gya} and UTfit~\cite{Bona:2006sa}.
This may look rather tolerant, in view of the recent progress~\cite{Bazavov:2016nty} 
in the estimation of the hadronic matrix element by lattice, which lead to  
$\sim 6$\% uncertainty in $M_{12}^{\rm SM}$~\cite{Aoki:2016frl}.
This is because the central values of $|M_{12}/M_{12}^{\rm SM}|$ are greater
than unity in Ref.~\cite{Charles:2015gya} and Ref.~\cite{Bona:2006sa}, while the $Z'$ contribution always enhances $|M_{12}|$ 
relative to the SM value under the assumption of a real-valued $g_{bs}^L$ with $g_{bs}^R = 0$.
On the other hand, a recent study~\cite{DiLuzio:2017fdq} finds the SM prediction 
of $\Delta m_{B_s^0} = 2|M_{12}|$ to be 1.8$\sigma$ above the measured value,
favoring $|M_{12}/M_{12}^{\rm SM}|$ smaller than unity.
This is opposite to the results by CKMfitter~\cite{Charles:2015gya} and UTfit~\cite{Bona:2006sa},
although both UTfit (Summer 2016 result) and Ref.~\cite{DiLuzio:2017fdq}
take into account the recent lattice result~\cite{Bazavov:2016nty}.
If the result of Ref.~\cite{DiLuzio:2017fdq} is the case, 
the $Z'$ contribution may enhance $|M_{12}|$ only up to $\sim1\%$ so that
$g_{bs}^L$ is strongly constrained. In this case the estimated signal significances at LHC 
would shrink down to insignificant values for $|g_{bb}^L| \lesssim |g_{bs}^L|$, unless a tiny RH $Z'bs$ coupling 
exists for the cancellation in $M_{12}$ and/or $g_{bs}^L$ is close to pure imaginary
so that it gives a negative contribution in $M_{12}$.
The latter implies a nearly imaginary $C_9^{\rm NP}$ and
would need a dedicated global analysis of $b \to s\ell\ell$ observables, 
as discussed in Ref.~\cite{DiLuzio:2017fdq}.
In any case, a consensus among the different groups seems to be still missing 
for the prediction of $M_{12}$ in the SM, and a better understanding would be required for its calculation. 
At the same time, improvements in lattice calculations and determinations of CKM parameters 
will also facilitate a more precise SM prediction for $M_{12}$.

Although we considered $Z'bb$ and $Z'bs$ couplings to be the only couplings to the quark sector, the
$Z'$ may also couple to other quarks in general. For instance, if a non zero $Z'cc$ coupling exists, 
the process $pp\to Z' c+X$ can be induced at LHC. Such a process can mimic 
the $pp\to Z' b+X$ signature if the final state $c$-jet gets misidentified as $b$-jet. 
This possibility can not be excluded yet as pointed out in Ref.~\cite{Hou:2018npi}, 
where a procedure to disentangle $pp\to Z' c+X$ and 
$pp\to Z' b+X$ is discussed with the simultaneous application of
both $c$- and $b$-tagging. We also remark that our estimation of the signal significances ignored  
various experimental uncertainties and the QCD corrections to the signal cross sections.

For illustration, we focused on $m_{Z'} = 200$ and 500 GeV. In general, heavier $Z'$
are possible. However, due to the fall in the parton luminosity with the resonance mass,
the achievable significances are lower than those of 200 GeV and 500 GeV 
in both the $pp\to Z'$ and $pp\to Z'b$ processes.

Our results illustrate three possible scenarios for the  LHC discovery and identification
of a $Z'$ that might be behind the $b \to s\ell\ell$ anomalies.
The first one is the case with the minimal assumption,
where the LH $Z'bs$ coupling is the only coupling to the quark sector.
In this case, the discovery of the $pp\to Z'+X \to \mu^+ \mu^- +X$ and $pp\to Z' b+X \to \mu^+ \mu^- b +X$ processes
may occur with the full HL-LHC data, but should be accompanied by
a $\sim 30$\% or larger enhancement in the $B_s$ mixing,
which can be tested following future improvements in the estimation of the $B_s$ mixing.
The second one is the case with a tiny but nonzero RH $Z'bs$ coupling
such that the $B_s$ mixing remains SM-like due to the cancellation of the $Z'$ effects.
In this case, the discovery of the two modes may occur with Run-3 data
(or perhaps even Run-2 data);
this scenario predicts a nonzero RH $b\to s$ current, with $C'_9 \sim 0.1 C_9^{\rm NP}$, 
which can be tested with improvements in $b \to s\ell\ell$ measurements 
by ATLAS, CMS, LHCb and Belle II. 
In particular, precise measurements of $R_K$ and $R_{K^*}$ by LHCb with Run-2 
or further dataset may pin down the chiral structure of the $b\to s$ current.
The third scenario is the case with a flavor-conserving $Z'bb$ coupling 
much larger than $Z'bs$.
In this case, the two modes may be discovered with Run-3 data without
a significant effect in the $B_s$ mixing and RH $b\to s$ current,
but the role of the observed resonance in $b \to s \ell\ell$ is obscured.
%%%%%%%%%%

\textbf{Note added:} While revising the manuscript we noticed that the CMS 13 TeV 
36 fb$^{-1}$ result~\cite{Sirunyan:2018exx} for a
heavy resonance search in the dilepton final state is now available. 
We find that the extracted upper limits~\cite{extrac2} on $g^L_{bb}$ from Ref.~\cite{Sirunyan:2018exx} 
are comparable to those from ATLAS~\cite{Aaboud:2017buh} and do not change the conclusion of our results.

\vspace{0.2cm}
\acknowledgements 
We thank Rahul Sinha and Arjun Menon for many discussions.
TM is tankful to Tanumoy Mandal for fruitful discussions.
MK is supported by grant MOST-106-2112-M-002-015-MY3,
and TM is supported by grant MOST 106-2811-M-002-187 of R.O.C. Taiwan.


\begin{thebibliography}{150}
\bibitem{Ammar:1993sh} 
  R.~Ammar {\it et al.} [CLEO Collaboration],
  %``Evidence for penguins: First observation of B ---> K* (892) gamma,''
  Phys.\ Rev.\ Lett.\  {\bf 71}, 674 (1993).
  %%CITATION = PRLTA,71,674;%%
  %647 citations counted in INSPIRE as of 24 Oct 2015

%\cite{DescotesGenon:2012zf}
\bibitem{DescotesGenon:2012zf} 
  S.~Descotes-Genon, J.~Matias, M.~Ramon and J.~Virto,
  %``Implications from clean observables for the binned analysis of $B -> K*\mu^+\mu^-$ at large recoil,''
  JHEP {\bf 1301}, 048 (2013)
  %doi:10.1007/JHEP01(2013)048
  [arXiv:1207.2753 [hep-ph]].
  %%CITATION = doi:10.1007/JHEP01(2013)048;%%
  %163 citations counted in INSPIRE as of 03 Nov 2017

\bibitem{Aaij:2013qta} 
  R.~Aaij {\it et al.} [LHCb Collaboration],
  %``Measurement of Form-Factor-Independent Observables in the Decay $B^{0} \to K^{*0} \mu^+ \mu^-$,''
  Phys.\ Rev.\ Lett.\  {\bf 111}, 191801 (2013)
  [arXiv:1308.1707 [hep-ex]].
  %%CITATION = ARXIV:1308.1707;%%
  %173 citations counted in INSPIRE as of 24 Oct 2015

\bibitem{Aaij:2015oid} 
  R.~Aaij {\it et al.} [LHCb Collaboration],
  %``Angular analysis of the $B^{0} \to K^{*0} \mu^{+} \mu^{-}$ decay using 3 fb$^{-1}$ of integrated luminosity,''
  JHEP {\bf 1602}, 104 (2016)
  %doi:10.1007/JHEP02(2016)104
  [arXiv:1512.04442 [hep-ex]].
  %%CITATION = doi:10.1007/JHEP02(2016)104;%%
  %195 citations counted in INSPIRE as of 11 Aug 2017

%\cite{Aaij:2014ora}
\bibitem{Aaij:2014ora} 
  R.~Aaij {\it et al.} [LHCb Collaboration],
  %``Test of lepton universality using $B^{+}\rightarrow K^{+}\ell^{+}\ell^{-}$ decays,''
  Phys.\ Rev.\ Lett.\  {\bf 113}, 151601 (2014)
%  doi:10.1103/PhysRevLett.113.151601
  [arXiv:1406.6482 [hep-ex]].
%  %%CITATION = doi:10.1103/PhysRevLett.113.151601;%%
  %350 citations counted in INSPIRE as of 12 May 2017

%\cite{Aaij:2017vbb}
\bibitem{Aaij:2017vbb} 
  R.~Aaij {\it et al.} [LHCb Collaboration],
  %``Test of lepton universality with $B^{0} \rightarrow K^{*0}\ell^{+}\ell^{-}$ decays,''
  JHEP {\bf 1708}, 055 (2017)
  [arXiv:1705.05802 [hep-ex]].
  %%CITATION = doi:10.1007/JHEP08(2017)055;%%
  %118 citations counted in INSPIRE as of 17 Jan 2018

\bibitem{Aaij:2014pli} 
  R.~Aaij {\it et al.} [LHCb Collaboration],
  %``Differential branching fractions and isospin asymmetries of $B \to K^{(*)} \mu^+ \mu^-$ decays,''
  JHEP {\bf 1406}, 133 (2014)
  [arXiv:1403.8044 [hep-ex]].
  %%CITATION = ARXIV:1403.8044;%%
  %60 citations counted in INSPIRE as of 24 Oct 2015

\bibitem{Aaij:2016flj} 
  R.~Aaij {\it et al.} [LHCb Collaboration],
  %``Measurements of the S-wave fraction in $B^{0}\rightarrow K^{+}\pi^{-}\mu^{+}\mu^{-}$ decays and the $B^{0}\rightarrow K^{\ast}(892)^{0}\mu^{+}\mu^{-}$ differential branching fraction,''
  JHEP {\bf 1611}, 047 (2016)
  %doi:10.1007/JHEP11(2016)047, 10.1007/JHEP04(2017)142, 10.1007/JHEP11(2016)047, 10.1007/JHEP04(2017)142
  [arXiv:1606.04731 [hep-ex]].
  %%CITATION = doi:10.1007/JHEP11(2016)047, 10.1007/JHEP04(2017)142, 10.1007/JHEP11(2016)047, 10.1007/JHEP04(2017)142;%%
  %37 citations counted in INSPIRE as of 11 Aug 2017


\bibitem{Aaij:2015esa} 
  R.~Aaij {\it et al.} [LHCb Collaboration],
  %``Angular analysis and differential branching fraction of the decay $B^0_s\to\phi\mu^+\mu^-$,''
  JHEP {\bf 1509}, 179 (2015)
  %doi:10.1007/JHEP09(2015)179
  [arXiv:1506.08777 [hep-ex]].
  %%CITATION = doi:10.1007/JHEP09(2015)179;%%
  %115 citations counted in INSPIRE as of 11 Aug 2017

\bibitem{Aaij:2015xza} 
  R.~Aaij {\it et al.} [LHCb Collaboration],
  %``Differential branching fraction and angular analysis of $\Lambda^{0}_{b} \rightarrow \Lambda \mu^+\mu^-$ decays,''
  JHEP {\bf 1506}, 115 (2015)
  %doi:10.1007/JHEP06(2015)115
  [arXiv:1503.07138 [hep-ex]].
  %%CITATION = doi:10.1007/JHEP06(2015)115;%%
  %44 citations counted in INSPIRE as of 11 Aug 2017

\bibitem{ATLAS:2017dlm} 
  ATLAS Collaboration,
  %``Angular analysis of $B^0_d \to K^{*}\mu^+\mu^-$ decays in $pp$ collisions at $\sqrt{s}= 8$ TeV with the ATLAS detector,''
  ATLAS-CONF-2017-023.
  %%CITATION = ATLAS-CONF-2017-023;%%


\bibitem{Sirunyan:2017dhj} 
  A.~M.~Sirunyan {\it et al.} [CMS Collaboration],
  %``Measurement of angular parameters from the decay $\mathrm{B}^0 \to \mathrm{K}^{*0} \mu^+ \mu^-$ in proton-proton collisions at $\sqrt{s} = $ 8 TeV,''
  arXiv:1710.02846 [hep-ex].
  %%CITATION = ARXIV:1710.02846;%%

%\cite{Wehle:2016yoi}
\bibitem{Wehle:2016yoi} 
  S.~Wehle {\it et al.} [Belle Collaboration],
  %``Lepton-Flavor-Dependent Angular Analysis of $B\to K^\ast \ell^+\ell^-$,''
  %Phys.\ Rev.\ Lett.\  {\bf 118}, no. 11, 111801 (2017)
  Phys.\ Rev.\ Lett.\  {\bf 118}, 111801 (2017)
%  doi:10.1103/PhysRevLett.118.111801
  [arXiv:1612.05014 [hep-ex]].
%  %%CITATION = doi:10.1103/PhysRevLett.118.111801;%%

\bibitem{Abe:2010gxa} 
  T.~Abe {\it et al.} [Belle-II Collaboration],
  %``Belle II Technical Design Report,''
  arXiv:1011.0352 [physics.ins-det].
  %%CITATION = ARXIV:1011.0352;%%
  %270 citations counted in INSPIRE as of 28 Oct 2015

% recent b -> s fits



% post R_K* b -> s fits

\bibitem{Capdevila:2017bsm} 
  B.~Capdevila, A.~Crivellin, S.~Descotes-Genon, J.~Matias and J.~Virto,
  %``Patterns of New Physics in $b\to s\ell^+\ell^-$ transitions in the light of recent data,''
  JHEP {\bf 1801}, 093 (2018)
  [arXiv:1704.05340 [hep-ph]].
  %%CITATION = ARXIV:1704.05340;%%
  %72 citations counted in INSPIRE as of 08 Nov 2017
  
\bibitem{Altmannshofer:2017yso} 
  W.~Altmannshofer, P.~Stangl and D.~M.~Straub,
  %``Interpreting Hints for Lepton Flavor Universality Violation,''
  %Phys.\ Rev.\ D {\bf 96}, no. 5, 055008 (2017)
  Phys.\ Rev.\ D {\bf 96}, 055008 (2017)
  %doi:10.1103/PhysRevD.96.055008
  [arXiv:1704.05435 [hep-ph]].
  %%CITATION = doi:10.1103/PhysRevD.96.055008;%%
  %58 citations counted in INSPIRE as of 08 Nov 2017

\bibitem{DAmico:2017mtc} 
  G.~D'Amico, M.~Nardecchia, P.~Panci, F.~Sannino, A.~Strumia, R.~Torre and A.~Urbano,
  %``Flavour anomalies after the $R_{K^*}$ measurement,''
  JHEP {\bf 1709}, 010 (2017)
  %doi:10.1007/JHEP09(2017)010
  [arXiv:1704.05438 [hep-ph]].
  %%CITATION = doi:10.1007/JHEP09(2017)010;%%
  %60 citations counted in INSPIRE as of 08 Nov 2017

\bibitem{Hiller:2017bzc} 
  G.~Hiller and I.~Nisandzic,
  %``$R_K$ and $R_{K^{\ast}}$ beyond the standard model,''
  %Phys.\ Rev.\ D {\bf 96}, no. 3, 035003 (2017)
  Phys.\ Rev.\ D {\bf 96}, 035003 (2017)
  %doi:10.1103/PhysRevD.96.035003
  [arXiv:1704.05444 [hep-ph]].
  %%CITATION = doi:10.1103/PhysRevD.96.035003;%%
  %48 citations counted in INSPIRE as of 08 Nov 2017

\bibitem{Geng:2017svp} 
  L.-S.~Geng, B.~Grinstein, S.~J\"ager, J.~Martin Camalich, X.-L.~Ren and R.-X.~Shi,
  %``Towards the discovery of new physics with lepton-universality ratios of $b\to s\ell\ell$ decays,''
  Phys.\ Rev.\ D {\bf 96}, no. 9, 093006 (2017)
  [arXiv:1704.05446 [hep-ph]].
  %%CITATION = ARXIV:1704.05446;%%
  %56 citations counted in INSPIRE as of 08 Nov 2017

\bibitem{Ciuchini:2017mik} 
  M.~Ciuchini, A.~M.~Coutinho, M.~Fedele, E.~Franco, A.~Paul, L.~Silvestrini and M.~Valli,
  %``On Flavourful Easter eggs for New Physics hunger and Lepton Flavour Universality violation,''
  %Eur.\ Phys.\ J.\ C {\bf 77}, no. 10, 688 (2017)
  Eur.\ Phys.\ J.\ C {\bf 77}, 688 (2017)
  %doi:10.1140/epjc/s10052-017-5270-2
  [arXiv:1704.05447 [hep-ph]].
  %%CITATION = doi:10.1140/epjc/s10052-017-5270-2;%%
  %49 citations counted in INSPIRE as of 08 Nov 2017

\bibitem{Celis:2017doq} 
  A.~Celis, J.~Fuentes-Martin, A.~Vicente and J.~Virto,
  %``Gauge-invariant implications of the LHCb measurements on lepton-flavor nonuniversality,''
  %Phys.\ Rev.\ D {\bf 96}, no. 3, 035026 (2017)
  Phys.\ Rev.\ D {\bf 96}, 035026 (2017)
  %doi:10.1103/PhysRevD.96.035026
  [arXiv:1704.05672 [hep-ph]].
  %%CITATION = doi:10.1103/PhysRevD.96.035026;%%
  %30 citations counted in INSPIRE as of 08 Nov 2017

%\cite{Hurth:2017hxg}
\bibitem{Hurth:2017hxg} 
  T.~Hurth, F.~Mahmoudi, D.~Martinez Santos and S.~Neshatpour,
  %``Lepton nonuniversality in exclusive $b{\rightarrow}s{\ell}{\ell}$ decays,''
  Phys.\ Rev.\ D {\bf 96}, no. 9, 095034 (2017)
  [arXiv:1705.06274 [hep-ph]].
  %%CITATION = doi:10.1103/PhysRevD.96.095034;%%
  %17 citations counted in INSPIRE as of 17 Jan 2018



\bibitem{Karan:2016wvu} 
  A.~Karan, R.~Mandal, A.~K.~Nayak, R.~Sinha and T.~E.~Browder,
  %``Signal of right-handed currents using $B\to K^*\ell^+\ell^-$ observables at the kinematic endpoint,''
  %Phys.\ Rev.\ D {\bf 95}, no. 11, 114006 (2017)
  Phys.\ Rev.\ D {\bf 95}, 114006 (2017)
  %doi:10.1103/PhysRevD.95.114006
  [arXiv:1603.04355 [hep-ph]].
  %%CITATION = doi:10.1103/PhysRevD.95.114006;%%


  
%\cite{Lees:2013kla}

% Z' models

%\cite{Altmannshofer:2013foa}
\bibitem{Altmannshofer:2013foa} 
  W.~Altmannshofer and D.~M.~Straub,
  %``New physics in $B \to K^*\mu\mu$?,''
  Eur.\ Phys.\ J.\ C {\bf 73}, 2646 (2013)
%  doi:10.1140/epjc/s10052-013-2646-9
  [arXiv:1308.1501 [hep-ph]].
%  %%CITATION = doi:10.1140/epjc/s10052-013-2646-9;%%
  %149 citations counted in INSPIRE as of 19 May 2016

%\cite{Gauld:2013qba}
\bibitem{Gauld:2013qba} 
  R.~Gauld, F.~Goertz and U.~Haisch,
  %``On minimal $Z'$ explanations of the $B\to K^*\mu^+\mu^-$ anomaly,''
  Phys.\ Rev.\ D {\bf 89}, 015005 (2014)
%  doi:10.1103/PhysRevD.89.015005
  [arXiv:1308.1959 [hep-ph]].
%  %%CITATION = doi:10.1103/PhysRevD.89.015005;%%
  %82 citations counted in INSPIRE as of 19 May 2016

%\cite{Buras:2013qja}
\bibitem{Buras:2013qja} 
  A.~J.~Buras and J.~Girrbach,
  %``Left-handed $Z'$ and $Z$ FCNC quark couplings facing new $b \to s \mu^+ \mu^-$ data,''
  JHEP {\bf 1312}, 009 (2013)
%  doi:10.1007/JHEP12(2013)009
  [arXiv:1309.2466 [hep-ph]].
%  %%CITATION = doi:10.1007/JHEP12(2013)009;%%
  %81 citations counted in INSPIRE as of 19 May 2016


%\cite{Gauld:2013qja}
\bibitem{Gauld:2013qja} 
  R.~Gauld, F.~Goertz and U.~Haisch,
  %``An explicit Z'-boson explanation of the $B \to K^* \mu^+ \mu^-$ anomaly,''
  JHEP {\bf 1401}, 069 (2014)
%  doi:10.1007/JHEP01(2014)069
  [arXiv:1310.1082 [hep-ph]].
%  %%CITATION = doi:10.1007/JHEP01(2014)069;%%
  %82 citations counted in INSPIRE as of 19 May 2016

%\cite{Buras:2013dea}
\bibitem{Buras:2013dea} 
  A.~J.~Buras, F.~De Fazio and J.~Girrbach,
  %``331 models facing new $b \to s\mu^+ \mu^-$ data,''
  JHEP {\bf 1402}, 112 (2014)
%  doi:10.1007/JHEP02(2014)112
  [arXiv:1311.6729 [hep-ph]].
%  %%CITATION = doi:10.1007/JHEP02(2014)112;%%
  %95 citations counted in INSPIRE as of 19 May 2016

%\cite{Ko:2013zsa}
\bibitem{Ko:2013zsa} 
  P.~Ko, Y.~Omura and C.~Yu,
  %``Higgs phenomenology in Type-I 2HDM with $U(1)_H$ Higgs gauge symmetry,''
  JHEP {\bf 1401}, 016 (2014)
%  doi:10.1007/JHEP01(2014)016
  [arXiv:1309.7156 [hep-ph]].
%  %%CITATION = doi:10.1007/JHEP01(2014)016;%%
  %16 citations counted in INSPIRE as of 24 Jul 2016

%\cite{Ahmed:2014vqa}
\bibitem{Ahmed:2014vqa} 
  I.~Ahmed, M.~J.~Aslam and M.~A.~Paracha,
  %``Effects of neutral $Z^{\prime}$ boson in $B_{s} \to \phi \ell^{+} \ell^{-}$ decay with polarized $\phi$ and the unpolarized and polarized $CP$ violation asymmetry,''
%  Phys.\ Rev.\ D {\bf 89}, no. 1, 015006 (2014)
  Phys.\ Rev.\ D {\bf 89}, 015006 (2014)
%  doi:10.1103/PhysRevD.89.015006
  [arXiv:1401.2162 [hep-ph]].
%  %%CITATION = doi:10.1103/PhysRevD.89.015006;%%
  %2 citations counted in INSPIRE as of 19 May 201

%\cite{Altmannshofer:2014cfa}
\bibitem{Altmannshofer:2014cfa} 
  W.~Altmannshofer, S.~Gori, M.~Pospelov and I.~Yavin,
  %``Quark flavor transitions in $L_\mu-L_\tau$ models,''
  Phys.\ Rev.\ D {\bf 89}, 095033 (2014)
%  doi:10.1103/PhysRevD.89.095033
  [arXiv:1403.1269 [hep-ph]].
%  %%CITATION = doi:10.1103/PhysRevD.89.095033;%%
  %101 citations counted in INSPIRE as of 19 May 2016  

  
%\cite{Bhattacharya:2014wla}
\bibitem{Bhattacharya:2014wla} 
  B.~Bhattacharya, A.~Datta, D.~London and S.~Shivashankara,
  %``Simultaneous Explanation of the $R_K$ and $R(D^{(*)})$ Puzzles,''
  Phys.\ Lett.\ B {\bf 742}, 370 (2015)
%  doi:10.1016/j.physletb.2015.02.011
  [arXiv:1412.7164 [hep-ph]].
%  %%CITATION = doi:10.1016/j.physletb.2015.02.011;%%
  %48 citations counted in INSPIRE as of 19 May 2016

%\cite{Crivellin:2015mga}
\bibitem{Crivellin:2015mga} 
  A.~Crivellin, G.~D'Ambrosio and J.~Heeck,
  %``Explaining $h\to\mu^\pm\tau^\mp$, $B\to K^* \mu^+\mu^-$ and $B\to K \mu^+\mu^-/B\to K e^+e^-$ in a two-Higgs-doublet model with gauged $L_\mu-L_\tau$,''
  Phys.\ Rev.\ Lett.\  {\bf 114}, 151801 (2015)
%  doi:10.1103/PhysRevLett.114.151801
  [arXiv:1501.00993 [hep-ph]].
%  %%CITATION = doi:10.1103/PhysRevLett.114.151801;%%
  %116 citations counted in INSPIRE as of 19 May 2016

%\cite{Crivellin:2015lwa}
\bibitem{Crivellin:2015lwa} 
  A.~Crivellin, G.~D'Ambrosio and J.~Heeck,
  %``Addressing the LHC flavor anomalies with horizontal gauge symmetries,''
%  Phys.\ Rev.\ D {\bf 91}, no. 7, 075006 (2015)
  Phys.\ Rev.\ D {\bf 91}, 075006 (2015)
%  doi:10.1103/PhysRevD.91.075006
  [arXiv:1503.03477 [hep-ph]].
%  %%CITATION = doi:10.1103/PhysRevD.91.075006;%%
  %79 citations counted in INSPIRE as of 19 May 2016

%\cite{Niehoff:2015bfa}
\bibitem{Niehoff:2015bfa} 
  C.~Niehoff, P.~Stangl and D.~M.~Straub,
  %``Violation of lepton flavour universality in composite Higgs models,''
  Phys.\ Lett.\ B {\bf 747}, 182 (2015)
%  doi:10.1016/j.physletb.2015.05.063
  [arXiv:1503.03865 [hep-ph]].
%  %%CITATION = doi:10.1016/j.physletb.2015.05.063;%%
  %32 citations counted in INSPIRE as of 19 May 2016

%\cite{Sierra:2015fma}
\bibitem{Sierra:2015fma} 
  D.~Aristizabal Sierra, F.~Staub and A.~Vicente,
  %``Shedding light on the $b\to s$ anomalies with a dark sector,''
%  Phys.\ Rev.\ D {\bf 92}, no. 1, 015001 (2015)
  Phys.\ Rev.\ D {\bf 92}, 015001 (2015)
%  doi:10.1103/PhysRevD.92.015001
  [arXiv:1503.06077 [hep-ph]].
%  %%CITATION = doi:10.1103/PhysRevD.92.015001;%%
  %44 citations counted in INSPIRE as of 19 May 2016


%\cite{Altmannshofer:2015sma}
\bibitem{Altmannshofer:2015sma} 
  W.~Altmannshofer and D.~M.~Straub,
  %``Implications of $b\to s$ measurements,''
  arXiv:1503.06199 [hep-ph].
  %%CITATION = ARXIV:1503.06199;%%
  %33 citations counted in INSPIRE as of 27 Oct 2015


%\cite{Crivellin:2015era}
\bibitem{Crivellin:2015era} 
  A.~Crivellin, L.~Hofer, J.~Matias, U.~Nierste, S.~Pokorski and J.~Rosiek,
  %``Lepton-flavour violating $B$ decays in generic $Z'$ models,''
%  Phys.\ Rev.\ D {\bf 92}, no. 5, 054013 (2015)
   Phys.\ Rev.\ D {\bf 92}, 054013 (2015)
%  doi:10.1103/PhysRevD.92.054013
  [arXiv:1504.07928 [hep-ph]].
%  %%CITATION = doi:10.1103/PhysRevD.92.054013;%%
  %33 citations counted in INSPIRE as of 04 May 2016

%\cite{Celis:2015ara}
\bibitem{Celis:2015ara} 
  A.~Celis, J.~Fuentes-Martin, M.~Jung and H.~Serodio,
  %``Family nonuniversal Z′ models with protected flavor-changing interactions,''
%  Phys.\ Rev.\ D {\bf 92}, no. 1, 015007 (2015)
  Phys.\ Rev.\ D {\bf 92}, 015007 (2015)
  [arXiv:1505.03079 [hep-ph]].
  %%CITATION = ARXIV:1505.03079;%%
  %4 citations counted in INSPIRE as of 03 Aug 2015
  
%\cite{Belanger:2015nma}
\bibitem{Belanger:2015nma} 
%  G.~Bélanger, C.~Delaunay and S.~Westhoff,
  G.~B\'elanger, C.~Delaunay and S.~Westhoff,
  %``A Dark Matter Relic From Muon Anomalies,''
  Phys.\ Rev.\ D {\bf 92}, 055021 (2015)
%  doi:10.1103/PhysRevD.92.055021
  [arXiv:1507.06660 [hep-ph]].
%  %%CITATION = doi:10.1103/PhysRevD.92.055021;%%
  %38 citations counted in INSPIRE as of 12 May 2017

%\cite{Falkowski:2015zwa}
\bibitem{Falkowski:2015zwa} 
  A.~Falkowski, M.~Nardecchia and R.~Ziegler,
  %``Lepton Flavor Non-Universality in B-meson Decays from a U(2) Flavor Model,''
  JHEP {\bf 1511}, 173 (2015)
%  doi:10.1007/JHEP11(2015)173
  [arXiv:1509.01249 [hep-ph]].
%  %%CITATION = doi:10.1007/JHEP11(2015)173;%%
  %12 citations counted in INSPIRE as of 04 May 2016

%\cite{Descotes-Genon:2015uva}
\bibitem{Descotes-Genon:2015uva} 
  S.~Descotes-Genon, L.~Hofer, J.~Matias and J.~Virto,
  %``Global analysis of $b\to s\ell\ell$ anomalies,''
  JHEP {\bf 1606}, 092 (2016)
  [arXiv:1510.04239 [hep-ph]].
  %%CITATION = ARXIV:1510.04239;%%
  %31 citations counted in INSPIRE as of 02 May 2016


%\cite{Allanach:2015gkd}
\bibitem{Allanach:2015gkd} 
  B.~Allanach, F.~S.~Queiroz, A.~Strumia and S.~Sun,
  %``$Z′$ models for the LHCb and $g-2$ muon anomalies,''
  %Phys.\ Rev.\ D {\bf 93}, no. 5, 055045 (2016)
  Phys.\ Rev.\ D {\bf 93}, 055045 (2016)
%  doi:10.1103/PhysRevD.93.055045
  [arXiv:1511.07447 [hep-ph]].
%  %%CITATION = doi:10.1103/PhysRevD.93.055045;%%
  %10 citations counted in INSPIRE as of 04 May 2016


%\cite{Buras:2015kwd}
\bibitem{Buras:2015kwd} 
  A.~J.~Buras and F.~De Fazio,
  %``$\varepsilon'/\varepsilon$ in 331 Models,''
  JHEP {\bf 1603}, 010 (2016)
%  doi:10.1007/JHEP03(2016)010
  [arXiv:1512.02869 [hep-ph]].
%  %%CITATION = doi:10.1007/JHEP03(2016)010;%%
  %7 citations counted in INSPIRE as of 04 May 2016

%\cite{Fuyuto:2015gmk}
\bibitem{Fuyuto:2015gmk} 
  K.~Fuyuto, W.-S.~Hou and M.~Kohda,
  %``Z′ -induced FCNC decays of top, beauty, and strange quarks,''
  %Phys.\ Rev.\ D {\bf 93}, no. 5, 054021 (2016)
  Phys.\ Rev.\ D {\bf 93}, 054021 (2016)
%  doi:10.1103/PhysRevD.93.054021
  [arXiv:1512.09026 [hep-ph]].
%  %%CITATION = doi:10.1103/PhysRevD.93.054021;%%
  %1 citations counted in INSPIRE as of 04 May 2016

%\cite{Chiang:2016qov}
\bibitem{Chiang:2016qov} 
  C.-W.~Chiang, X.-G.~He and G.~Valencia,
  %``Z′ model for b→sℓ$\overline{ℓ}$ flavor anomalies,''
  %Phys.\ Rev.\ D {\bf 93}, no. 7, 074003 (2016)
  Phys.\ Rev.\ D {\bf 93}, 074003 (2016)
%  doi:10.1103/PhysRevD.93.074003
  [arXiv:1601.07328 [hep-ph]].
%  %%CITATION = doi:10.1103/PhysRevD.93.074003;%%
  %4 citations counted in INSPIRE as of 04 May 2016

\bibitem{Kim:2016bdu} 
  C.S.~Kim, X.-B.~Yuan and Y.-J.~Zheng,
  %``Constraints on a Z′ boson within minimal flavor violation,''
  %Phys.\ Rev.\ D {\bf 93}, no. 9, 095009 (2016)
  Phys.\ Rev.\ D {\bf 93}, 095009 (2016)
  %doi:10.1103/PhysRevD.93.095009
  [arXiv:1602.08107 [hep-ph]].
  %%CITATION = doi:10.1103/PhysRevD.93.095009;%%

%\cite{Altmannshofer:2016oaq}
\bibitem{Altmannshofer:2016oaq} 
  W.~Altmannshofer, M.~Carena and A.~Crivellin,
  %``A $L_\mu - L_\tau$ theory of Higgs flavor violation and $(g-2)_\mu$,''
  Phys.\ Rev.\ D {\bf 94}, no. 9, 095026 (2016)
  [arXiv:1604.08221 [hep-ph]].
  %%CITATION = ARXIV:1604.08221;%%


%\cite{Hisano:2016afc}
\bibitem{Hisano:2016afc} 
  J.~Hisano, Y.~Muramatsu, Y.~Omura and Y.~Shigekami,
  %``Flavor physics induced by light $Z'$ from SO(10) GUT,''
  JHEP {\bf 1611}, 018 (2016)
  [arXiv:1607.05437 [hep-ph]].
  %%CITATION = ARXIV:1607.05437;%%

\bibitem{Altmannshofer:2016brv} 
  W.~Altmannshofer, C.-Y.~Chen, P.~S.~Bhupal Dev and A.~Soni,
  %``Lepton flavor violating Z′ explanation of the muon anomalous magnetic moment,''
  Phys.\ Lett.\ B {\bf 762}, 389 (2016)
  %doi:10.1016/j.physletb.2016.09.046
  [arXiv:1607.06832 [hep-ph]].
  %%CITATION = doi:10.1016/j.physletb.2016.09.046;%%
  %25 citations counted in INSPIRE as of 29 Nov 2017

%\cite{Ko:2017quv}
\bibitem{Ko:2017quv} 
  P.~Ko, T.~Nomura and H.~Okada,
  %``A flavor dependent gauge symmetry, Predictive radiative seesaw and LHCb anomalies,''
  Phys.\ Lett.\ B {\bf 772}, 547 (2017)
  [arXiv:1701.05788 [hep-ph]].
  %%CITATION = ARXIV:1701.05788;%%
  %10 citations counted in INSPIRE as of 12 May 2017

%\cite{Bhatia:2017tgo}
\bibitem{Bhatia:2017tgo} 
  D.~Bhatia, S.~Chakraborty and A.~Dighe,
  %``Neutrino mixing and $R_K$ anomaly in U(1)$_X$ models: a bottom-up approach,''
  JHEP {\bf 1703}, 117 (2017)
%  doi:10.1007/JHEP03(2017)117
  [arXiv:1701.05825 [hep-ph]].
%  %%CITATION = doi:10.1007/JHEP03(2017)117;%%
  %6 citations counted in INSPIRE as of 12 May 2017  

\bibitem{Hou:2017ozb} 
  W.-S.~Hou, M.~Kohda and T.~Modak,
  %``Search for $tZ'$ associated production induced by $tcZ'$ couplings at the LHC,''
  %Phys.\ Rev.\ D {\bf 96}, no. 1, 015037 (2017)
  Phys.\ Rev.\ D {\bf 96}, 015037 (2017)
  %doi:10.1103/PhysRevD.96.015037
  [arXiv:1702.07275 [hep-ph]].
  %%CITATION = doi:10.1103/PhysRevD.96.015037;%%
  
\bibitem{Alok:2017jgr}
  A.~K.~Alok, B.~Bhattacharya, D.~Kumar, J.~Kumar, D.~London and S.~U.~Sankar,
  %``New physics in $b \rightarrow s \mu^+ \mu^-$: Distinguishing models through CP-violating effects,''
  Phys.\ Rev.\ D {\bf 96}, no. 1, 015034 (2017)
  [arXiv:1703.09247 [hep-ph]].
  %%CITATION = doi:10.1103/PhysRevD.96.015034;%%
  
  
   
\bibitem{DiChiara:2017cjq} 
  S.~Di Chiara, A.~Fowlie, S.~Fraser, C.~Marzo, L.~Marzola, M.~Raidal and C.~Spethmann,
  %``Minimal flavor-changing $Z'$ models and muon $g-2$ after the $R_{K^*}$ measurement,''
  Nucl.\ Phys.\ B {\bf 923}, 245 (2017)
  %doi:10.1016/j.nuclphysb.2017.08.003
  [arXiv:1704.06200 [hep-ph]].
  %%CITATION = doi:10.1016/j.nuclphysb.2017.08.003;%%

%\cite{Greljo:2017vvb}
\bibitem{Greljo:2017vvb} 
  A.~Greljo and D.~Marzocca,
  %``High-$p_T$ dilepton tails and flavour physics,''
  Eur.\ Phys.\ J.\ C {\bf 77}, no. 8, 548 (2017)
  [arXiv:1704.09015 [hep-ph]].
  %%CITATION = ARXIV:1704.09015;%%
  %2 citations counted in INSPIRE as of 12 May 2017  
  
 
  
  
\bibitem{Alok:2017sui}
  A.~K.~Alok, B.~Bhattacharya, A.~Datta, D.~Kumar, J.~Kumar and D.~London,
  %``New Physics in $b \to s \mu^+ \mu^-$ after the Measurement of $R_{K^*}$,''
  Phys.\ Rev.\ D {\bf 96}, no. 9, 095009 (2017)
  [arXiv:1704.07397 [hep-ph]].
  %%CITATION = doi:10.1103/PhysRevD.96.095009;%%   
  
%\cite{Bonilla:2017lsq}
\bibitem{Bonilla:2017lsq} 
  C.~Bonilla, T.~Modak, R.~Srivastava and J.~W.~F.~Valle,
  %``$U(1)_{B_3-3L_\mu}$ gauge symmetry as the simplest description of $b\to s$ anomalies,''
  arXiv:1705.00915 [hep-ph].
  %%CITATION = ARXIV:1705.00915;%%
  %2 citations counted in INSPIRE as of 12 May 2017 
  
%\cite{Ellis:2017nrp}
\bibitem{Ellis:2017nrp} 
  J.~Ellis, M.~Fairbairn and P.~Tunney,
  %``Anomaly-Free Models for Flavour Anomalies,''
  arXiv:1705.03447 [hep-ph].
  %%CITATION = ARXIV:1705.03447;%%
  %1 citations counted in INSPIRE as of 12 May 2017  
  
 %\cite{Bishara:2017pje}
\bibitem{Bishara:2017pje} 
  F.~Bishara, U.~Haisch and P.~F.~Monni,
  %``On Light Resonance Interpretations of the B Decay Anomalies,''
  Phys.\ Rev.\ D {\bf 96}, no. 5, 055002 (2017)
  [arXiv:1705.03465 [hep-ph]].
  %%CITATION = ARXIV:1705.03465;%% 
  
%\cite{Alonso:2017uky}
\bibitem{Alonso:2017uky} 
  R.~Alonso, P.~Cox, C.~Han and T.~T.~Yanagida,
  %``Flavoured $B-L$ Local Symmetry and Anomalous Rare $B$ Decays,''
  Phys.\ Lett.\ B {\bf 774}, 643 (2017)
  [arXiv:1705.03858 [hep-ph]].
  %%CITATION = ARXIV:1705.03858;%%   
  
\bibitem{Chiang:2017hlj} 
  C.-W.~Chiang, X.-G.~He, J.~Tandean and X.-B.~Yuan,
  %``$R_{K^{(*)}}$ and related $b\to s\ell\bar\ell$ anomalies in minimal flavor violation framework with $Z'$ boson,''
  %Phys.\ Rev.\ D {\bf 96}, no. 11, 115022 (2017)
  Phys.\ Rev.\ D {\bf 96}, 115022 (2017)
  %doi:10.1103/PhysRevD.96.115022
  [arXiv:1706.02696 [hep-ph]].
  %%CITATION = doi:10.1103/PhysRevD.96.115022;%%
  
  
%\cite{King:2017anf}
\bibitem{King:2017anf} 
  S.~F.~King,
  %``Flavourful $Z'$ models for $R_{K^{(*)}}$,''
  JHEP {\bf 1708}, 019 (2017)
  [arXiv:1706.06100 [hep-ph]].
  %%CITATION = ARXIV:1706.06100;%%
  
%\cite{Chivukula:2017qsi}
\bibitem{Chivukula:2017qsi} 
  R.~S.~Chivukula, J.~Isaacson, K.~A.~Mohan, D.~Sengupta and E.~H.~Simmons,
  %``$R_K$ anomalies and simplified limits on $Z'$ models at the LHC,''
  Phys.\ Rev.\ D {\bf 96}, no. 7, 075012 (2017)
  [arXiv:1706.06575 [hep-ph]].
  %%CITATION = ARXIV:1706.06575;%%   
  
  
%\cite{Cline:2017ihf}
\bibitem{Cline:2017ihf} 
  J.~M.~Cline and J.~Martin Camalich,
  %``$B$ decay anomalies from nonabelian local horizontal symmetry,''
  Phys.\ Rev.\ D {\bf 96}, no. 5, 055036 (2017)
  [arXiv:1706.08510 [hep-ph]].
  %%CITATION = doi:10.1103/PhysRevD.96.055036;%%
  %12 citations counted in INSPIRE as of 25 Mar 2018  
  
%\cite{Cox:2017eme}
\bibitem{Cox:2017eme} 
  P.~Cox, C.~Han and T.~T.~Yanagida,
  %``LHC Search for Right-handed Neutrinos in $Z^\prime$ Models,''
  JHEP {\bf 1801}, 037 (2018)
  [arXiv:1707.04532 [hep-ph]].
  %%CITATION = ARXIV:1707.04532;%%
  %2 citations counted in INSPIRE as of 30 Oct 2017
  
  
  %\cite{Baek:2017sew}
\bibitem{Baek:2017sew} 
  S.~Baek,[106]
  %``Dark matter contribution to $b\to s \mu^+ \mu^-$ anomaly in local $U(1)_{L_\mu-L_\tau}$ model,''
  arXiv:1707.04573 [hep-ph].
  %%CITATION = ARXIV:1707.04573;exactrc%%
  %1 citations counted in INSPIRE as of 30 Oct 2017
  

%\cite{Romao:2017qnu}
\bibitem{Romao:2017qnu} 
  M.~C.~Romao, S.~F.~King and G.~K.~Leontaris,
  %``Non-universal $Z'$ from Fluxed GUTs,''
  arXiv:1710.02349 [hep-ph].
  %%CITATION = ARXIV:1710.02349;%%

  %\cite{Cox:2017rgn}
\bibitem{Cox:2017rgn} 
  P.~Cox, C.~Han and T.~T.~Yanagida,
  %``Right-handed Neutrino Dark Matter in a U(1) Extension of the Standard Model,''
  JCAP {\bf 1801}, no. 01, 029 (2018)
  [arXiv:1710.01585 [hep-ph]].
  %%CITATION = ARXIV:1710.01585;%% 

\bibitem{Faisel:2017glo} 
  G.~Faisel and J.~Tandean,
  %``Connecting $b\to s\ell\bar\ell$ anomalies to enhanced rare nonleptonic $\bar{B}{}_s^0$ decays in $Z'$ model,''
  JHEP {\bf 1802}, 074 (2018)
  [arXiv:1710.11102 [hep-ph]].
  %%CITATION = ARXIV:1710.11102;%%  

\bibitem{Dalchenko:2017shg} 
  M.~Dalchenko, B.~Dutta, R.~Eusebi, P.~Huang, T.~Kamon and D.~Rathjens,
  %``Bottom-quark Fusion Processes at the LHC for Probing $Z^{\prime}$ Models and B-meson Decay Anomalies,''
  arXiv:1707.07016 [hep-ph].
  %%CITATION = ARXIV:1707.07016;%% 

%\cite{Allanach:2017bta}
\bibitem{Allanach:2017bta} 
  B.~C.~Allanach, B.~Gripaios and T.~You,
  %``The case for future hadron colliders from $B \to K^{(*)} \mu^+ \mu^-$  decays,''
  JHEP {\bf 1803}, 021 (2018)
  [arXiv:1710.06363 [hep-ph]].
  %%CITATION = doi:10.1007/JHEP03(2018)021;%%
  %5 citations counted in INSPIRE as of 19 Mar 2018

\bibitem{Choudhury:2017ijp} 
  D.~Choudhury, A.~Kundu, R.~Mandal and R.~Sinha,
  %``$R_{K^{(*)}}$ and $R(D^{(*)})$ anomalies resolved with lepton mixing,''
  arXiv:1712.01593 [hep-ph].
  %%CITATION = ARXIV:1712.01593;%%

\bibitem{Antusch:2017tud} 
  S.~Antusch, C.~Hohl, S.~F.~King and V.~Susic,
  %``Non-universal Z' from SO(10) GUTs with vector-like family and the origin of neutrino masses,''
  arXiv:1712.05366 [hep-ph].

\bibitem{Fuyuto:2017sys} 
  K.~Fuyuto, H.-L.~Li and J.-H.~Yu,
  %``Implications of hidden gauged $U(1)$ model for $B$ anomalies,''
  arXiv:1712.06736 [hep-ph].

\bibitem{Raby:2017igl} 
  S.~Raby and A.~Trautner,
  %``A "Vector-like chiral" fourth family to explain muon anomalies,''
  arXiv:1712.09360 [hep-ph].

\bibitem{DiLuzio:2017fdq} 
  L.~Di Luzio, M.~Kirk and A.~Lenz,
  %``One constraint to kill them all?,''
  arXiv:1712.06572 [hep-ph].
  %%CITATION = ARXIV:1712.06572;%%
  
  
%\cite{Chala:2018igk}
\bibitem{Chala:2018igk} 
  M.~Chala and M.~Spannowsky,
  %``On the behaviour of composite resonances breaking lepton flavour universality,''
  arXiv:1803.02364 [hep-ph].
  %%CITATION = ARXIV:1803.02364;%%  

%%%
% Z' references should be updated?
%%%

% end of Z' models

\bibitem{Buras:2001ra} 
  A.~J.~Buras, S.~J\"ager and J.~Urban,
  %``Master formulae for Delta F=2 NLO QCD factors in the standard model and beyond,''
  Nucl.\ Phys.\ B {\bf 605}, 600 (2001)
  %doi:10.1016/S0550-3213(01)00207-3
  [hep-ph/0102316].
  %%CITATION = doi:10.1016/S0550-3213(01)00207-3;%%

\bibitem{Olive:2016xmw} 
  C.~Patrignani {\it et al.} [Particle Data Group],
  %``Review of Particle Physics,''
  %Chin.\ Phys.\ C {\bf 40}, no. 10, 100001 (2016).
  Chin.\ Phys.\ C {\bf 40}, 100001 (2016) and 2017 update.
  %doi:10.1088/1674-1137/40/10/100001
  %%CITATION = doi:10.1088/1674-1137/40/10/100001;%%
  %1337 citations counted in INSPIRE as of 11 Aug 2017

\bibitem{Capdevila:2016ivx} 
  B.~Capdevila, S.~Descotes-Genon, J.~Matias and J.~Virto,
  %``Assessing lepton-flavour non-universality from $B\to K^*\ell\ell$ angular analyses,''
  JHEP {\bf 1610}, 075 (2016)
%  doi:10.1007/JHEP10(2016)075
  [arXiv:1605.03156 [hep-ph]].
%  %%CITATION = doi:10.1007/JHEP10(2016)075;%%



\bibitem{Lenz:2010gu} 
  A.~Lenz {\it et al.},
  %``Anatomy of New Physics in $B - \bar{B}$ mixing,''
  Phys.\ Rev.\ D {\bf 83}, 036004 (2011)
%  doi:10.1103/PhysRevD.83.036004
  [arXiv:1008.1593 [hep-ph]].
%  %%CITATION = doi:10.1103/PhysRevD.83.036004;%%

\bibitem{Charles:2015gya} 
  J.~Charles {\it et al.},
  %``Current status of the Standard Model CKM fit and constraints on $\Delta F=2$ New Physics,''
  %Phys.\ Rev.\ D {\bf 91}, no. 7, 073007 (2015)[106]
  Phys.\ Rev.\ D {\bf 91}, 073007 (2015)
  %doi:10.1103/PhysRevD.91.073007
  [arXiv:1501.05013 [hep-ph]].
  %%CITATION = doi:10.1103/PhysRevD.91.073007;%%

\bibitem{Bona:2006sa} 
  M.~Bona {\it et al.} [UTfit Collaboration],
  %``Constraints on new physics from the quark mixing unitarity triangle,''
  Phys.\ Rev.\ Lett.\  {\bf 97}, 151803 (2006)
  %doi:10.1103/PhysRevLett.97.151803
  [hep-ph/0605213];
  %%CITATION = doi:10.1103/PhysRevLett.97.151803;%%
%\bibitem{Bona:2007vi} 
  %M.~Bona {\it et al.} [UTfit Collaboration],
  %``Model-independent constraints on $\Delta F=2$ operators and the scale of new physics,''
  JHEP {\bf 0803}, 049 (2008)
  %doi:10.1088/1126-6708/2008/03/049
  [arXiv:0707.0636 [hep-ph]];
  %%CITATION = doi:10.1088/1126-6708/2008/03/049;%%
  \url{http://www.utfit.org}, for updates.

\bibitem{Aoki:2016frl} 
  S.~Aoki {\it et al.},
  %``Review of lattice results concerning low-energy particle physics,''
  %Eur.\ Phys.\ J.\ C {\bf 77}, no. 2, 112 (2017)
  Eur.\ Phys.\ J.\ C {\bf 77}, 112 (2017)
  %doi:10.1140/epjc/s10052-016-4509-7
  [arXiv:1607.00299 [hep-lat]];
  %%CITATION = doi:10.1140/epjc/s10052-016-4509-7;%%
  \url{http://flag.unibe.ch}, for updates.

\bibitem{Bazavov:2016nty}
  A.~Bazavov {\it et al.} [Fermilab Lattice and MILC Collaborations],
  %``$B^0_{(s)}$-mixing matrix elements from lattice QCD for the Standard Model and beyond,''
  %Phys.\ Rev.\ D {\bf 93} (2016) no.11,  113016
  Phys.\ Rev.\ D {\bf 93} (2016),  113016
%  doi:10.1103/PhysRevD.93.113016
  [arXiv:1602.03560 [hep-lat]].
%  %%CITATION = doi:10.1103/PhysRevD.93.113016;%%
  %51 citations counted in INSPIRE as of 30 Jul 2017



\bibitem{Bona-ICHEP2016}
 For a projection of the $B_s$ mixing constraint, see, e.g. the talk by 
 M. Bona presented at ICHEP2016, Chicago, USA, August 2016;
\url{https://indico.cern.ch/event/432527/contributions/1071767/}.

% b -> snunu

\bibitem{Buras:2014fpa} 
  A.~J.~Buras, J.~Girrbach-Noe, C.~Niehoff and D.~M.~Straub,
  %``$ B\to {K}^{\left(\ast \right)}\nu \overline{\nu} $ decays in the Standard Model and beyond,''
  JHEP {\bf 1502}, 184 (2015)
%  doi:10.1007/JHEP02(2015)184
  [arXiv:1409.4557 [hep-ph]].
%  %%CITATION = doi:10.1007/JHEP02(2015)184;%%

\bibitem{Brod:2010hi} 
  J.~Brod, M.~Gorbahn and E.~Stamou,
  %``Two-Loop Electroweak Corrections for the $K \to \pi \nu \bar{\nu}$ Decays,''
  Phys.\ Rev.\ D {\bf 83}, 034030 (2011)
%  doi:10.1103/PhysRevD.83.034030
  [arXiv:1009.0947 [hep-ph]].
%  %%CITATION = doi:10.1103/PhysRevD.83.034030;%%
  %214 citations counted in INSPIRE as of 09 May 2017

\bibitem{Amhis:2016xyh} 
  Y.~Amhis {\it et al.},
  %``Averages of $b$-hadron, $c$-hadron, and $\tau$-lepton properties as of summer 2016,''
  Eur.\ Phys.\ J.\ C {\bf 77}, no. 12, 895 (2017)
  [arXiv:1612.07233 [hep-ex]]
  %%CITATION = ARXIV:1612.07233;%%
  and online update at \url{http://www.slac.stanford.edu/xorg/hflav}.


\bibitem{Grygier:2017tzo} 
  J.~Grygier {\it et al.} [Belle Collaboration],
  %``Search for $\boldsymbol{B\to h\nu\bar{\nu}}$ decays with semileptonic tagging at Belle,''
  Phys.\ Rev.\ D {\bf 96}, no. 9, 091101 (2017)
  [arXiv:1702.03224 [hep-ex]].
  %%CITATION = ARXIV:1702.03224;%%

%

\bibitem{Aaboud:2017buh} 
  M.~Aaboud {\it et al.} [ATLAS Collaboration],
  JHEP {\bf 1710}, 182 (2017)
  [arXiv:1707.02424 [hep-ex]].
  The $\sigma(pp\to Z'+ X)\cdot \mathcal{B}(Z'\to\mu^+\mu^-)$ 95\% CL upper limit is 
  available, along with other auxiliaries, at 
  \url{https://atlas.web.cern.ch/Atlas/GROUPS/PHYSICS/PAPERS/EXOT-2016-05/}.
%  which is not displayed in the published version.



\bibitem{extrac}
We digitize figure for dimuon 95\% CL
upper limit obtained from \url{https://atlas.web.cern.ch/Atlas/GROUPS/PHYSICS/PAPERS/EXOT-2016-05/}
to extract the $\sigma(pp\to Z'+ X)\cdot \mathcal{B}(Z'\to\mu^+\mu^-)$ upper limit for pole masses 
$m_{Z'} = 200$ and 500 GeV.


\bibitem{Alwall:2014hca} 
  J.~Alwall {\it et al.},
  %``The automated computation of tree-level and next-to-leading order differential cross sections, and their matching to parton shower simulations,''
  JHEP {\bf 1407}, 079 (2014)
%  doi:10.1007/JHEP07(2014)079
  [arXiv:1405.0301 [hep-ph]].
%  %%CITATION = doi:10.1007/JHEP07(2014)079;%%
  %1396 citations counted in INSPIRE as of 08 Nov 2016


\bibitem{Ball:2013hta} 
  R.~D.~Ball {\it et al.} [NNPDF Collaboration],
  %``Parton distributions with QED corrections,''
  Nucl.\ Phys.\ B {\bf 877}, 290 (2013)
  %doi:10.1016/j.nuclphysb.2013.10.010
  [arXiv:1308.0598 [hep-ph]].
  %%CITATION = doi:10.1016/j.nuclphysb.2013.10.010;%%

\bibitem{Altmannshofer:2014pba} 
  W.~Altmannshofer, S.~Gori, M.~Pospelov and I.~Yavin,
  %``Neutrino Trident Production: A Powerful Probe of New Physics with Neutrino Beams,''
  Phys.\ Rev.\ Lett.\  {\bf 113}, 091801 (2014)
  %doi:10.1103/PhysRevLett.113.091801
  [arXiv:1406.2332 [hep-ph]].
  %%CITATION = doi:10.1103/PhysRevLett.113.091801;%%
  %100 citations counted in INSPIRE as of 28 Nov 2017

\bibitem{Mishra:1991bv} 
  S.~R.~Mishra {\it et al.} [CCFR Collaboration],
  %``Neutrino tridents and W Z interference,''
  Phys.\ Rev.\ Lett.\  {\bf 66}, 3117 (1991).
%  doi:10.1103/PhysRevLett.66.3117
%  %%CITATION = doi:10.1103/PhysRevLett.66.3117;%%
  %65 citations counted in INSPIRE as of 09 May 2017

%\cite{Haisch:2011up}
\bibitem{Haisch:2011up} 
  U.~Haisch and S.~Westhoff,
  %``Massive Color-Octet Bosons: Bounds on Effects in Top-Quark Pair Production,''
  JHEP {\bf 1108}, 088 (2011)
%  doi:10.1007/JHEP08(2011)088
  [arXiv:1106.0529 [hep-ph]].
%  %%CITATION = doi:10.1007/JHEP08(2011)088;%%
  %76 citations counted in INSPIRE as of 09 May 2017



%\cite{ALEPH:2005ab}
\bibitem{ALEPH:2005ab} 
  S.~Schael {\it et al.} 
  %[ALEPH and DELPHI and L3 and OPAL and SLD Collaborations and LEP Electroweak Working Group and SLD Electroweak Group and SLD Heavy Flavour Group],
   [ALEPH, DELPHI, L3, OPAL and SLD Collaborations, and LEP Electroweak Working 
   Group, SLD Electroweak Group and SLD Heavy Flavour Group],
  %``Precision electroweak measurements on the $Z$ resonance,''
  Phys.\ Rept.\  {\bf 427}, 257 (2006)
%  doi:10.1016/j.physrep.2005.12.006
  [hep-ex/0509008].
%  %%CITATION = doi:10.1016/j.physrep.2005.12.006;%%
  %1585 citations counted in INSPIRE as of 09 May 2017 

\bibitem{Fuster:1999dj} 
  J.~Fuster {\it et al.} [DELPHI Collaboration],
  %``A Search for FCNC in Z --> b anti-d, b anti-s decays,''
  CERN-OPEN-99-393, CERN-DELPHI-99-81.
  %%CITATION = CERN-OPEN-99-393, CERN-DELPHI-99-81;%%

%\cite{Pospelov:2008zw}
\bibitem{Pospelov:2008zw} 
  M.~Pospelov,
  %``Secluded U(1) below the weak scale,''
  Phys.\ Rev.\ D {\bf 80}, 095002 (2009)
  [arXiv:0811.1030 [hep-ph]].
%  %%CITATION = doi:10.1103/PhysRevD.80.095002;%%
  %431 citations counted in INSPIRE as of 08 Jul 2017
  
  
%\cite{Jegerlehner:2009ry}
\bibitem{Jegerlehner:2009ry} 
  F.~Jegerlehner and A.~Nyffeler,
  %``The Muon g-2,''
  Phys.\ Rept.\  {\bf 477}, 1 (2009)
  [arXiv:0902.3360 [hep-ph]].
%  %%CITATION = doi:10.1016/j.physrep.2009.04.003;%%
  %672 citations counted in INSPIRE as of 08 Jul 2017

\bibitem{Aaij:2017vad} 
  R.~Aaij {\it et al.} [LHCb Collaboration],
  %``Measurement of the $B^0_s\to\mu^+\mu^-$ branching fraction and effective lifetime and search for $B^0\to\mu^+\mu^-$ decays,''
  %Phys.\ Rev.\ Lett.\  {\bf 118}, no. 19, 191801 (2017)
  Phys.\ Rev.\ Lett.\  {\bf 118}, 191801 (2017)
  %doi:10.1103/PhysRevLett.118.191801
  [arXiv:1703.05747 [hep-ex]].
  %%CITATION = doi:10.1103/PhysRevLett.118.191801;%%
  %60 citations counted in INSPIRE as of 04 Dec 2017
 

\bibitem{Alloul:2013bka} 
  A.~Alloul, N.~D.~Christensen, C.~Degrande, C.~Duhr and B.~Fuks,
  %``FeynRules  2.0 - A complete toolbox for tree-level phenomenology,''
  Comput.\ Phys.\ Commun.\  {\bf 185}, 2250 (2014)
%  doi:10.1016/j.cpc.2014.04.012
  [arXiv:1310.1921 [hep-ph]].
%  %%CITATION = doi:10.1016/j.cpc.2014.04.012;%%
  %517 citations counted in INSPIRE as of 09 Dec 2016   
  
\bibitem{Sjostrand:2006za}
  T.~Sj\" ostrand, S.~Mrenna and P.~Skands,
  %``PYTHIA 6.4 Physics and Manual,''
  JHEP {\bf 0605}, 026 (2006).
  %doi:10.1088/1126-6708/2006/05/026
  %[hep-ph/0603175].
  %%CITATION = doi:10.1088/1126-6708/2006/05/026;%%
  
%\cite{Alwall:2007fs}
\bibitem{Alwall:2007fs} 
  J.~Alwall {\it et al.},
  %``Comparative study of various algorithms for the merging of parton showers and matrix elements in hadronic collisions,''
  Eur.\ Phys.\ J.\ C {\bf 53}, 473 (2008)
%  %doi:10.1140/epjc/s10052-007-0490-5
  [arXiv:0706.2569 [hep-ph]].
%  %%CITATION = doi:10.1140/epjc/s10052-007-0490-5;%%
  %555 citations counted in INSPIRE as of 24 Feb 2017  
  

%\cite{deFavereau:2013fsa}
\bibitem{deFavereau:2013fsa} 
  J.~de Favereau {\it et al.}  [DELPHES 3 Collaboration],
  %``DELPHES 3, A modular framework for fast simulation of a generic collider experiment,''
  JHEP {\bf 1402}, 057 (2014)
  [arXiv:1307.6346 [hep-ex]].
  %%CITATION = ARXIV:1307.6346;%%
  %177 citations counted in INSPIRE as of 14 Feb 2015

 %\cite{Li:2012wna}
\bibitem{Li:2012wna} 
  Y.~Li and F.~Petriello,
  %``Combining QCD and electroweak corrections to dilepton production in FEWZ,''
  Phys.\ Rev.\ D {\bf 86}, 094034 (2012)
%  doi:10.1103/PhysRevD.86.094034
  [arXiv:1208.5967 [hep-ph]].


\bibitem{twiki}
  ATLAS-CMS recommended predictions for top-quark-pair cross sections:
  \url{https://twiki.cern.ch/twiki/bin/view/LHCPhysics/TtbarNNLO}.
  

%\cite{Kidonakis:2010ux}
\bibitem{Kidonakis:2010ux} 
  N.~Kidonakis,
  %``Two-loop soft anomalous dimensions for single top quark associated production with a W- or H-,''
  Phys.\ Rev.\ D {\bf 82}, 054018 (2010)
  [arXiv:1005.4451 [hep-ph]].
%  %%CITATION = doi:10.1103/PhysRevD.82.054018;%%
  %433 citations counted in INSPIRE as of 21 janv. 2016
  
%\cite{Gehrmann:2014fva}
\bibitem{Gehrmann:2014fva} 
  %T.~Gehrmann, M.~Grazzini, S.~Kallweit, P.~Maierhöfer, A.~von Manteuffel, S.~Pozzorini, D.~Rathlev and L.~Tancredi,
  T.~Gehrmann, M.~Grazzini, S.~Kallweit, P.~Maierh\"ofer, A.~von Manteuffel,
  S.~Pozzorini, D.~Rathlev and L.~Tancredi,
  %``$W^+W^-$ Production at Hadron Colliders in Next to Next to Leading Order QCD,''
  %Phys.\ Rev.\ Lett.\  {\bf 113}, no. 21, 212001 (2014)
  Phys.\ Rev.\ Lett.\  {\bf 113}, 212001 (2014)
%  doi:10.1103/PhysRevLett.113.212001
  [arXiv:1408.5243 [hep-ph]].
%  %%CITATION = doi:10.1103/PhysRevLett.113.212001;%%
  %124 citations counted in INSPIRE as of 26 Nov 2016  
  
  %\cite{Grazzini:2016swo}
\bibitem{Grazzini:2016swo} 
  M.~Grazzini, S.~Kallweit, D.~Rathlev and M.~Wiesemann,
  %``$W^{\pm}Z$ production at hadron colliders in NNLO QCD,''
  Phys.\ Lett.\ B {\bf 761}, 179 (2016)
%  doi:10.1016/j.physletb.2016.08.017
  [arXiv:1604.08576 [hep-ph]].
%  %%CITATION = doi:10.1016/j.physletb.2016.08.017;%%
  %22 citations counted in INSPIRE as of 21 Oct 2016  
  
%\cite{Cascioli:2014yka}
\bibitem{Cascioli:2014yka} 
  F.~Cascioli {\it et al.},
  %``ZZ production at hadron colliders in NNLO QCD,''
  Phys.\ Lett.\ B {\bf 735}, 311 (2014)
%  doi:10.1016/j.physletb.2014.06.056
  [arXiv:1405.2219 [hep-ph]].
%  %%CITATION = doi:10.1016/j.physletb.2014.06.056;%%
  %118 citations counted in INSPIRE as of 26 Nov 2016    

%\cite{Cowan:2010js}
\bibitem{Cowan:2010js} 
  G.~Cowan, K.~Cranmer, E.~Gross and O.~Vitells,
  %``Asymptotic formulae for likelihood-based tests of new physics,''
  Eur.\ Phys.\ J.\ C {\bf 71}, 1554 (2011)
  Erratum: [Eur.\ Phys.\ J.\ C {\bf 73}, 2501 (2013)]
%  doi:10.1140/epjc/s10052-011-1554-0, 10.1140/epjc/s10052-013-2501-z
  [arXiv:1007.1727 [physics.data-an]].
%  %%CITATION = doi:10.1140/epjc/s10052-011-1554-0, 10.1140/epjc/s10052-013-2501-z;%%
  %1446 citations counted in INSPIRE as of 28 Jul 2017

%\cite{Boughezal:2016isb}
\bibitem{Boughezal:2016isb} 
  R.~Boughezal, X.~Liu and F.~Petriello,
  %``Phenomenology of the Z-boson plus jet process at NNLO,''
  Phys.\ Rev.\ D {\bf 94}, no. 7, 074015 (2016)
  [arXiv:1602.08140 [hep-ph]].
  %%CITATION = ARXIV:1602.08140;%%
  

 %\cite{ATLAS:2014ffa}
\bibitem{ATLAS:2014ffa} 
  ATLAS Collaboration,
  %``Estimation of non-prompt and fake lepton backgrounds in final states with top quarks produced in proton-proton collisions at \sqrt{s}=8~TeV with the ATLAS detector,''
  ATLAS-CONF-2014-058.
  %%CITATION = ATLAS-CONF-2014-058;%%
  %41 citations counted in INSPIRE as of 22 Nov 2016

% 

\bibitem{Buras:2012jb} 
  A.~J.~Buras, F.~De Fazio and J.~Girrbach,
  %``The Anatomy of Z' and Z with Flavour Changing Neutral Currents in the Flavour Precision Era,''
  JHEP {\bf 1302}, 116 (2013)
  %doi:10.1007/JHEP02(2013)116
  [arXiv:1211.1896 [hep-ph]].
  %%CITATION = doi:10.1007/JHEP02(2013)116;%%
  
  
%\cite{Hou:2018npi}
\bibitem{Hou:2018npi} 
  W.-S.~Hou, M.~Kohda and T.~Modak,
  %``Unraveling the couplings of a Drell-Yan produced $Z'$ with heavy-flavor tagging,''
  arXiv:1801.02579 [hep-ph].
  %%CITATION = ARXIV:1801.02579;%%
 

%\cite{Sirunyan:2018exx}
\bibitem{Sirunyan:2018exx} 
  A.~M.~Sirunyan {\it et al.} [CMS Collaboration],
  %``Search for high-mass resonances in dilepton final states in proton-proton collisions at $\sqrt{s}=$ 13 TeV,''
  arXiv:1803.06292 [hep-ex].
  %%CITATION = ARXIV:1803.06292;%%
  %1 citations counted in INSPIRE as of 30 Mar 2018 
  %
  
\bibitem{extrac2}
The CMS analysis puts 95\% CL upper limits on the quantity $R_\sigma$, 
which is defined as the ratio of dimuon production cross section via $Z'$ to the one 
via $Z$ or $\gamma^*$ (in the dimuon-invariant mass range of 60--120 GeV):
$
R_\sigma = \frac{ \sigma(pp\to Z'+ X\to\mu^+\mu^- + X) }{ \sigma(pp\to Z + X\to\mu^+\mu^- + X) }. 
$
As in the case of ATLAS limit, we digitize the figure for the dimuon final state to extract the limit on $R_\sigma$.
If the $Z'$ width is narrow, the $R_\sigma$ can be interpreted as the limits on $\sigma(pp\to Z'+ X)\cdot \mathcal{B}(Z'\to\mu^+\mu^-)$,
with the  multiplication of the SM prediction of $\sigma(pp\to Z + X)\mathcal{B}(Z\to\mu^+\mu^-)=1928.0$ 
pb. See Ref.\cite{Hou:2017ozb} for the details of how CMS limits are extracted.

\end{thebibliography}
\end{document}